\documentclass[12pt]{article}
\usepackage{amsmath,amssymb,epsfig}
\usepackage{cancel}
\usepackage{color}
\usepackage{caption}

\makeatletter

\@addtoreset{equation}{section}
\makeatother

\newcommand{\red}{\textcolor{black}}
\newcommand{\blue}{\textcolor{black}}
\newcommand{\redtwo}{\textcolor{black}}

\begin{document}

\title{\vbox{
\baselineskip 14pt
\hfill \hbox{\normalsize KUNS-2515, EPHOU-14-015}
}
\vskip 1.7cm
\bf \red{Standard Model-like D-brane models  \\ and gauge couplings}\vskip 0.5cm
}
\author{
Yuta~Hamada$^{1}$, \ \
Tatsuo~Kobayashi$^{2}$, \ and \
Shohei~Uemura$^{1}$
\\*[20pt]
{\it \normalsize 
${}^{1}$Department of Physics, Kyoto University, 
Kyoto 606-8502, Japan}
\\
{\it \normalsize 
${}^{2}$Department of Physics, Hokkaido University, 
Sapporo, 060-0810 Japan}
\\*[50pt]}

\date{
\centerline{\small \bf Abstract}
\begin{minipage}{0.9\linewidth}
\medskip 
\medskip 
\small
We systematically search 
intersecting D-brane models, which  just realize the Standard Model chiral matter contents and gauge symmetry.
We construct new classes of non-supersymmetric Standard Model-like models.
We also study the gauge coupling constants of these models.
The tree level gauge coupling is a function of the compactification moduli, the string scale, the string coupling and the winding numbers of D-branes.
By tuning them, we examine whether 
the models can explain
the experimental values of gauge couplings.
As a result, we find that the string scale should be greater than $10^{14-15}$GeV if the compactification scale and the string scale are of the same order.
\end{minipage}
}

\newpage

\begin{titlepage}
\maketitle
\thispagestyle{empty}
\clearpage
\end{titlepage}

\renewcommand{\thefootnote}{\arabic{footnote}}
\setcounter{footnote}{0}

\section{Introduction}

The Standard Model (SM) is one of the greatest achievements of particle physics.
It is consistent with all of the experimental results by tuning about 19 free parameters and succeeded in predicting new physics.
The discovery of the Higgs scalar \cite{Aad:2012tfa,Chatrchyan:2012ufa} is the latest example.
However,  many questions still remain in particle physics. 
What is the quantum theory of gravity?
How does the mysterious flavor structure of the SM appear? 
What is the origin of neutrino masses, inflation, dark matter and other cosmological observations?

From the viewpoint of quantum gravity, superstring theory is the most promising candidate to successfully describe it, and almost the only candidate available.
Furthermore, superstring theory is also a unified theory of other interactions and matter fields.
Superstring theory naturally has gauge symmetry.
There appear gravitons, gauge bosons, matter fermions, and scalars in its massless spectrum. 
Thus, it is important to construct stringy theories explaining the SM.

The intersecting D-brane models are an interesting technique to realize four-dimensional (4D) chiral gauge theories as low-energy effective theory from superstring theory \cite{Berkooz:1996km,Blumenhagen:2000wh,Aldazabal:2000dg,Angelantonj:2000hi,
Ibanez:2001nd} (for review, see \cite{Blumenhagen:2006ci,Ibanez} and references therein).
In these models, chiral matter fermions are realized as the R-sector of open strings stretching between D-branes at angles, while gauge bosons are realized as open strings on the same set of 
D-branes.
It is surprising that simple compactification models realize the SM spectrum or supersymmetric SM spectrum as zero modes.
For example, in \cite{Ibanez:2001nd}, the intersecting D-brane model with just the SM spectrum 
was constructed, which we call the IMR model in this paper.
Similarly,  supersymmetric SM\blue{-like} models were constructed (see e.g \cite{Cvetic:2001nr, Honecker:2004kb, Cremades:2002qm}).

In addition to the massless spectrum, it is quite important to explain the quantitative  structure of the SM, 
i.e. the gauge couplings, Yukawa couplings and the Higgs potential parameters as well as possibly neutrino 
Majorana masses.
\red{In this paper, we focus on the gauge couplings.}
In 4D low-energy effective theories derived from 
heterotic string theory, the gauge couplings at tree level are unified up to Kac-Moody levels $\kappa_a$
at the string scale \cite{Ginsparg:1987ee}, 
which is of ${\cal O}(10^{17})$ GeV \cite{Kaplunovsky:1987rp}.
This prediction is very strong.
In order to explain the experimental values, 
we may need some corrections, e.g. stringy threshold corrections \cite{Dixon:1990pc,Antoniadis:1991fh,Derendinger:1991hq}.
(See for numerical studies e.g. Refs.~\cite{Ibanez:1991zv,Kawabe:1994mj}.)

On the other hand, the gauge coupling is a function of the D-brane volume in D-brane models.
In intersecting D-brane models, gauge groups of the SM are originating from 
different D-branes, which have volumes independent of each other.
Thus, at first sight, it seems always possible to explain the three gauge couplings of the SM by tuning volume moduli, 
because the number of parameters, moduli, is sufficiently larger than three.\footnote{In Ref.~\cite{Blumenhagen:2003jy}, 
a specific relation among the three gauge couplings is shown in a certain class of supersymmetric models.}
However, in an explicit model, \red{the values of volume moduli} are constrained by other conditions.
For example, tachyonic modes may appear for some values of moduli 
in non-supersymmetric models.
Also, the string coupling $g_s$ may be required to be strong for some values of 
moduli to derive realistic values of the SM gauge couplings.
However, our theory is reliable at the weak string coupling.
Then, it is non-trivial to explain the three SM gauge couplings under the above conditions.

In this paper, we study systematically the model construction of intersecting D-brane models.
We construct  new classes of non-supersymmetric SM-like models, which have 
the same gauge symmetry and chiral matter contents as those of the SM but no exotics except right-handed neutrinos.
We show three classes of SM-like models.
We study their gauge couplings as well as those of the IMR model 
under the above constraints.

This paper is organized as follows.
In section 2, we briefly review the intersecting D-brane models.
In section 3, we construct new classes of SM-like models.
We  calculate gauge couplings in section 4.
Section 5 is our conclusion.
In Appendix A, we discuss the systematic search for SM-like models.
In Appendix B, we discuss one-loop threshold corrections due to massive modes.

\section{Intersecting D-brane model building}

In this section, we briefly review the toroidal orientifold models with intersecting D6-branes.
We first consider Type IIA superstring theory compactified on a factorized six-dimensional torus $T^{6}=T_{1}^{2}\times T_{2}^{2} \times T_{3}^{2}$ with intersecting D6-branes, 
where $T_{i}^{2}$ is the $i$-th two-dimensional torus; the two-dimensional Euclidean space modded by a lattice, 
\begin{equation}
T_{\blue{k}}^{2}={\bf C}/L(\tau_{\blue{k}}),
\nonumber
\end{equation}
\begin{equation}
L(\tau_{\blue{k}})=\{ z_{\blue{k}} \in {\bf C} | z_{\blue{k}} =\blue{m} i 
+ \blue{n} \tau_{\blue{k}}, 
\ n,m\in {\bf Z} \},
\end{equation}
where $\tau_{i}\in {\bf C}$.

$D6_a$-branes wrap 3-cycles $[\Pi_a]$ on $T^6$. 
Here, we restrict ourselves to the D-brane system in which
 all D6-brane's 3-cycles [$\Pi_{a}$] are factorized, $[\Pi_{a}]=[\Pi^{1}_{a}]\times [\Pi^{2}_{a}]\times [\Pi^{3}_{a}]$, where $[\Pi^{i}_{a}]$ is a 1-cycle of $T_{i}^{2}$. 
Then we can 
specify
the 3-cycles by using 6 integer winding numbers $(n_{a}^{i},m_{a}^{i})$.
$n_{a}^{i}$ is the winding number along the $\tau_{i}$ direction and $m_{a}^{i}$ is the winding number along the imaginary axis of $z_i$.
The intersection number between the D6$_{a}$-brane and the D6$_{b}$-brane  is denoted by $I_{ab}$
which
is determined by the winding numbers,
\begin{equation}
I_{ab}=[\Pi_{a}]\circ[\Pi_{b}]=\Pi_{i=1}^{3} \left( n_{a}^{i} m_{b}^{i} -m_{a}^{i} n_{b}^{i} \right).
\label{eq:intersection_number}
\end{equation}

The open string stretching between the D6$_{a}$-branes and the D6$_{b}$-branes has the \red{following} boundary conditions,
\begin{equation}{\rm Re} \frac{\partial}{\partial \sigma}e^{-i\theta_{a}^{i}}z_{i}|_{\sigma=0}=0,\ \  {\rm Im}\frac{d}{dt}e^{-i\theta_{a}^{i}}z_{i}|_{\sigma=0}=0, \end{equation}
\begin{equation}{\rm Re} \frac{\partial}{\partial \sigma}e^{-i\theta_{b}^{i}}z_{i} |_{\sigma=\pi}=0,\ \ {\rm Im}\frac{d}{dt}e^{-i\theta_{b}^{i}}z_{i} |_{\sigma=\pi}=0, \end{equation}
where 
\begin{equation}
\blue{\theta_{a}^{i} = {\rm tan}^{-1}\left( \frac{m_{a}^{i}+n_{a}^{i}{\rm Im}\tau_{i}}{n_{a}^{i}{\rm Re}\tau_{i}} \right)},
\end{equation}
is the angle of the D6$_{a}$-branes on the $i$-th torus.
These boundary conditions resolve the degeneracy of the ground states in the R-sector.
The resultant ground state corresponds to a 4D massless chiral fermion. 
Scalars appear in the NS-sector.
The ground state in the NS-sector depends on the intersecting angles $\theta_{ab}^{i}=(\theta_{b}^{i}-\theta_{a}^{i})/\pi$.
\blue{Assuming $1>\theta_{ab}^{i}>0$,}
the masses squared of four candidates for the lightest state are shown in Table \ref{tab:NS_light_spectrum}.
They would be massive, massless or tachyonic \red{depending on the angles}.
If there are massless states, a part of supersymmetry is recovered.
For example, when $\theta_{ba}^{1}+\theta_{ba}^{2}-\theta_{ba}^{3}=0$,
the first state in Table \ref{tab:NS_light_spectrum}
 is the massless ground state and the others are massive. 

\begin{table}[h]
\begin{center}
\begin{tabular}{cc}
\hline
\hline
State �& Mass$^2$\\
\hline
1 & $\frac{1}{\alpha'}(\theta_{ba}^{1}+\theta_{ba}^{2}-\theta_{ba}^{3})$\\ 
2 & $\frac{1}{\alpha'}(\theta_{ba}^{1}-\theta_{ba}^{2}+\theta_{ba}^{3})$\\ 
3 & $\frac{1}{\alpha'}(-\theta_{ba}^{1}+\theta_{ba}^{2}+\theta_{ba}^{3})$\\
4 & $\frac{1}{\alpha'}(1-\frac{1}{2}(\theta_{ba}^{1}+\theta_{ba}^{2}+\theta_{ba}^{3}))$\\ 
\hline
\hline
\end{tabular}
\end{center}
\caption{The masses squared of the light scalar states.}
\label{tab:NS_light_spectrum}
\end{table}

In this way, each intersection point has a 4D massless chiral fermion as well as scalars.
Also, a stack of $N_a$
D6$_a$-branes has gauge symmetry $U(N_a)$.
The open strings ending at the D6$_a$-branes have Chan-Paton charges, which correspond 
to the fundamental representation of $U(N_a)$.
This class of models leads to 
4D chiral $U(N)$ Yang-Mills theory as the low energy effective theory.
This fact is essential to derive the SM at low energy.

Now, we introduce the orientifold.\footnote{
We need the orientifold projection in order to obtain just the SM massless spectrum even if we do not consider supersymmetric models~\cite{Ibanez}.
}
The toroidal orientifold is obtained by modding $T^{6}$ by reflection operator $\mathcal{R}$, 
\begin{equation}
\mathcal{R}\ : \ {\rm Im}z_{1,2,3}  \rightarrow -{\rm Im}z_{1,2,3}.
\end{equation}
To define this operator $\mathcal{R}$ well, ${\rm Im}\tau_{i}$ in $L(\tau_{i})$ must be either 0 or 1/2.
The torus is rectangular for ${\rm Im}\tau_{i}=0$, while the torus is tilted for ${\rm Im}\tau_{i}=1/2$.
It is useful to define new ``winding numbers" $(\tilde{n}_{a}^{i}, \tilde{m}_{a}^{i})$, where $\tilde{n}_{a}^{i}=n_{a}^{i}$ and $\tilde{m}_{a}^{i}=m_{a}^{i}+{\rm Im}\tau_{i} n_{a}^{i}$. Hereafter, we use $(\tilde{n}_{a}^{i}, \tilde{m}_{a}^{i})$ as the winding numbers of a D6$_a$-brane on the $i$-th torus.

In this setup, we can construct perturbative vacua which have several stacks of $N_{a}$ D6$_{a}$-branes wrapping the whole 4D Minkowski spacetime and factorized 3-cycles $[\Pi_{a}]$ of $T^{6}$.
In addition to D6$_{a}$-branes, we need their orientifold mirror D6$_{a*}$-branes such that the system is $\mathcal{R}$-invariant.
The D6$_{a*}$-brane's winding numbers must be $(\tilde{n}_{a}^{i},-\tilde{m}_{a}^{i})$.

In the presence of an orientifold, the gauge symmetry $G_{a}$ appearing on D6$_a$-branes depends on whether the D6$_{a}$-branes lie on top of their orientifold mirror D$6_{a^*}$-branes or not.
If the D6$_{a}$-branes are apart from the D6$_{a^*}$-branes, the gauge group is $U(N_{a})$. Otherwise the gauge group is $Sp(2N_{a})$ or $SO(2N_{a})$.
The intersection points between D6$_{a}$-branes and D6$_{b}$-branes have massless 4D chiral fermions transforming as the bifundamental representation under $G_a \times G_b$. 
For example, if $G_{a,b}=U(N_{a,b})$, they transform as $(N_{a},\overline{N_{b}})$ under $U(N_a) \times U(N_b)$.

The number of intersection points $I_{ab}$ is obtained as
\begin{equation}
I_{ab}=\Pi_{i=1}^{3} \left( \tilde{n}_{a}^{i} \tilde{m}_{b}^{i} - \tilde{m}_{a}^{i} \tilde{n}_{b}^{i} \right).
\end{equation}

Using this D-brane system, we can realize a lot of patterns of chiral (super) Yang-Mills theories as effective theory, but not all patterns of theories.

\red{Next, let us discuss the constraints on intersecting D-brane models.}
D-branes have RR charges \red{which} must be canceled in compact space.
This constraint is derived from D-brane kinematics, and the same as Gauss's law of electromagnetism in compact space.
This is \red{called} the RR tadpole cancellation condition.
\red{Since} the RR charge is proportional to the D-brane homology, the constraint is written by 
\begin{equation}
\sum_{a=1,\cdots ,N} N_{a} [\Pi_{a}] - 4[\Pi_{O6}]=0,
\end{equation}
where $[\Pi_{O6}]$ is a cycle of the O6-planes.

In general, the gauge symmetry includes several $U(1)$ factors.
Some of them
become massive by the generalized Green-Schwartz mechanism.
That is,  $U(1)$ gauge bosons have non-zero couplings with RR-forms, especially $C_{5}$ and have non-perturbative St\"{u}ckelberg masses.
The coupling between $U(1)_{a}$ gauge boson and $C_{5}$ is obtained by the Chern-Simons term,
\begin{equation}
S_{CS}=\sum_{a} N_{a}\int_{D6_{a}}C_{5}\wedge {\rm tr}F_{a} +\cdots .
\label{eq:DbraneCS}
\end{equation}
We introduce $[\alpha_{k}]$ as the basis of 3-cycles and its dual basis $[\beta_{l}]$, where [$\alpha_k]\circ [\beta_l]=\delta_{kl}$.
We define 
\begin{equation}B_{2}^{k}=\int_{[\alpha_{k}]}C_{5}.
\end{equation}
Then the coupling between $U(1)$ gauge bosons and $B_2^k$ can be written by 
\begin{equation}S_{4D-CS}= N_{a}Q_{ak}\int_{M^{4}}B_{2}^{k}{\rm tr}F_{a}+ \cdots, 
\label{eq:4dFC}
\end{equation}
where \red{$Q_{ak}= [\Pi_a]\circ[\beta_k]$}.
This coupling induces masses of $U(1)$ gauge bosons.
The $U(1)$ gauge boson corresponding to 
$U(1)_{X}=\sum_{a} c_{a} U(1)_{a}$ is massless if and only if 
\red{
$\sum_{a} c_{a} \blue{N_{a}}[\Pi_{a}]\circ[\beta_{k}]\blue{-}\sum_{a*} c_{a} \blue{N_{a}}[\Pi_{a*}]\circ[\beta_{k}]=0$ 
}
for any $k$.
Otherwise, the $U(1)$ gauge boson becomes massive 
even if it is anomaly-free.

In the next section, 
we will construct intersecting D-brane models which have the same gauge group as \red{that of the} SM.
\red{
We will show that we can get the exact SM gauge group by using above mechanism to make extra gauge bosons massive.} 


\section{The SM-like models}

Our aim is to construct perturbative vacua which lead to SM-like effective theories by using type IIA orientifold.
For such a purpose, we systematically search vacua satisfying the following conditions:
\begin{itemize}
\item Gauge symmetry is the same as that of the SM 
\red{up to} the hidden sector, $SU(3)\times SU(2)\times U(1)_{Y} \times G_{hidden}$.
\item The chiral massless spectrum is the same as \red{that of} the SM with three right-handed neutrinos up to the hidden sector. 
\end{itemize}
For the RR tadpole cancellation, 
we need right-handed neutrinos  and the $G_{hidden}$ sector.
The matter fields in the hidden sector are  singlets under the SM gauge group.

There are two methods to realize the $SU(2)$ gauge symmetry.
One is to use a stack of two 
D6$_a$-branes separating from their orientifold mirror D6$_{a*}$-branes.
The theory in the worldvolume of the D6$_a$-brane is $U(2)$ Yang-Mills theory 
which contains $SU(2)$ group as subgroup.
We call this class of models $SU(2)$ models.
In this scenario, we must use a tilted torus to cancel the $U(2)$ anomaly.
There are many models using the $SU(2)$ method, see for the model satisfying the above condition, e.g. \cite{Ibanez:2001nd}.
The other is to use \red{one} D6$_{a}$-brane whose orientifold mirror D6$_{a*}$-brane is coincident with the D6$_{a}$-brane.
In this case, the gauge group can be enhanced from $U(1)$ to $Sp(2)$.
$Sp(2)$ is isomorphic to $SU(2)$ as Lie algebra.
Then, we can get the $SU(2)$ gauge symmetry.
We call this class of models $Sp(2)$ models.

We concentrate on the latter models in the following way:

\begin{itemize}
\item We construct
$Sp(2)$ models
where
$SU(2)$ gauge symmetry 
is realized by one brane and its orientifold mirror.
\end{itemize}

We can satisfy these conditions by using four stacks of branes, D6$_{a,b,c,d}$-branes. 
The multiplicity of the D6$_{a}$-branes N$_{a}$ is equal to three, and the others are one.
The D6$_{b}$-brane is on top of the O6-planes on one two-dimensional torus and perpendicular to them on the other two two-dimensional tori to realize $Sp(2)$ gauge symmetry.
The intersection numbers of these branes are required as follows,
\begin{equation} I_{ab}=3;\ I_{ac} = -3;\ I_{ac^*}=-3;\ I_{ad}=0;\ I_{ad^*} = 0,\nonumber \end{equation}
\begin{equation} I_{bc}=0;\ I_{db} = 3;\ I_{dc}= -3;\ I_{dc^*} = -3,\nonumber \end{equation}
\begin{equation} I_{aa^*}=0;\ I_{cc^*} = 0;\ I_{dd^*}=0,\label{eq:int_num}\end{equation}
such that the chiral spectrum of this model realizes the SM matter contents and realizes the gauge symmetry.
For the desired zero mode, we require the D6$_{a,c,d}$-branes to be parallel to the O-plane on at least one torus, too.
The hypercharge $U(1)_Y$ corresponds to the following linear combination of $U(1)$s,
\begin{equation}
U(1)_{Y}=\frac{1}{6}U(1)_{a} - \frac{1}{2} U(1)_{c} - \frac{1}{2} U(1)_{d}.
\end{equation}
There is some arbitrariness of the definition of $U(1)_{Y}$, but we can absorb it by renaming the branes.
In Table \ref{tab:matter_of_contents}, we summarize the chiral spectrum of this model, quantum numbers of non-Abelian and Abelian gauge symmetries, and their names in the SM.

\begin{table}[h]
\begin{center}
\begin{tabular}{ccccccc}
\hline
Intersection & name & $SU(3)\times SU(2)$ & $Q_{a}$ & $Q_{c}$ & $Q_{d}$ & Hypercharge \\
\hline
\hline
(ab) & $Q_{L}$ & 3(3,2)  &  1    & 0   & 0 & $\frac{1}{6}$ \\
(ac) &  $U_{R}$ & 3($\bar{3}$,1)  &  -1  & 1   & 0 & $-\frac{2}{3}$ \\
(ac$^{*}$) &  $D_{R}$ & 3($\bar{3}$,1) & -1 & -1 & 0 & $\frac{1}{3}$ \\
(db) &  $L$ & 3(1,2)       &  0    &  0  & 1 & $-\frac{1}{2}$ \\
(dc) &  $N_{R}$ & 3(1,1) &  0     &  1 & \redtwo{-}1 & 0 \\
(dc$^{*}$) & $E_{R}$  & 3(1,1) & 0  & -1 & -1 & 1 \\ 
\hline
\end{tabular}
\end{center}
\caption{Chiral matter contents. All the SM chiral fields appear in \redtwo{intersection} points as zero modes of the open string R sector.}
\label{tab:matter_of_contents}
\end{table}

We carry out a systematic analysis on all the possible D-brane configurations,
(see Appendix \ref{app:systematic} for the details).
As a result, it is found that general solutions realizing \red{Eq.\eqref{eq:int_num}} are classified into \blue{two} classes of models.

\begin{table}[h]
\begin{center}
\begin{tabular}{|c||c|c|c|}
\hline
 D-brane  & $T_{1}^{2}$ & $T_{2}^{2}$ & $T_{3}^{2}$\\
\hline
\hline
a &  ($\redtwo{1},0$)  &  ($\frac{a}{\beta_{2}}-1,\beta_{2}m_{a}^2$)  & ($-\frac{\redtwo{3 \epsilon}}{m_{a}^{2}},\beta_{3}m_{a}^{3}$)\\
b &  (0,$\redtwo{\epsilon_1}$)  &  ($\redtwo{\epsilon_{2}}/\beta_{2},0$)  & (0,$\redtwo{\epsilon \epsilon_1 \epsilon_{2}}$)\\
c &  ($\redtwo{-\beta_3 m_a^3 \left( n_a^2 +3\epsilon_5 n_d^2 \frac{m_a^2}{m_d^2} \right)},\redtwo{\epsilon_{3}}$)  &  ($\redtwo{\epsilon_{4}}/\beta_{2} ,0$)  & ($0,-\redtwo{\epsilon \epsilon_{3} \epsilon_{4}}$)\\
d &  ($\redtwo{\epsilon_5},0$)  &  ($\frac{d}{\beta_{2}}-1,\beta_{2}\redtwo{\epsilon_5}m_{d}^2$)  & ($\frac{\redtwo{3 \epsilon}}{m_{d}^{2}},-3\beta_{3}\frac{m_{a}^{2}}{m_{d}^{2}}m_{a}^{3}$)\\
\hline
\end{tabular}
\end{center}
\caption{\blue{General solutions of the $Sp(2)$ models.
$\beta_{2,3}=1-{\rm Im}\tau_{2,3}\in \{1,1/2\}$ and ${\rm Im}\tau_{1}$ is always zero. 
The $n,m$s are integer parameters and satisfy $m_{a}^{2},m_{d}^{2}$ are divisors
of 3 and $\epsilon_{i}$s are $\pm1$. $a,d,m_{a}^{3}$ are arbitrary integers and $n_{a,d}^{2}=\frac{a,d}{\beta_{2}}-1$.
}}
\label{tab:Sp(2)SM}
\end{table}

Both of them have the desired chiral spectrum.
However,  one of them can not make the extra $U(1)$ gauge boson massive through the Green-Schwartz mechanism while the $U(1)_Y$ gauge boson remains massless(see Appendix A).
This extra $U(1)$ symmetry corresponds to $U(1)_{B-L}$.
That is, both $U(1)_Y$  and $U(1)_{B-L}$ gauge bosons are massless or massive at the same time in that class of models.
The other  can make the $U(1)_{B-L}$ gauge boson massive with the $U(1)_Y$ gauge boson remaining massless.
Thus, this class of models can reproduce the SM chiral spectrum and gauge symmetry.
It is shown in Table \ref{tab:Sp(2)SM}.
There are no other solutions satisfying the conditions.
Note that gauginos and adjoint scalars appear in the gauge sector of our models
which would become massive by loop corrections~\cite{Ibanez:2001nd}.

For later calculation, we classify the models into three new further classes, as shown in Table \ref{tab:0til-SM}, Table \ref{tab:1til-SM} and Table \ref{tab:2til-SM}.
We refer to the class of models in Table \ref{tab:0til-SM} as 0til-SM, because they have no tilted torus. 
Also we refer
the class of models in Table \ref{tab:1til-SM} and Table \ref{tab:2til-SM} as 1til-SM and 2til-SM, respectively.
As we show in Table \ref{tab:Sp(2)SM}, we can not construct the SM-like models using three tilted tori since 
they always lead to an even number of generations.

\begin{table}[h]
\begin{center}
\begin{tabular}{|c||c|c|c|}
\hline
D-brane �& $T_{1}^{2}$ & $T_{2}^{2}$ & $T_{3}^{2}$\\
\hline
\hline
a &  ($\redtwo{1}$,0)  &  ($n_{a}^2,m_{a}^2$)  & ($-\frac{\redtwo{3 \epsilon}}{m_{a}^{2}},m_{a}^3$)\\
b &  (0,$\redtwo{\epsilon_1}$)  &  ($\redtwo{\epsilon_2},$0)  & (0,$\redtwo{\epsilon \epsilon_1 \epsilon_2}$)\\
c &  ($-\redtwo{\epsilon_3}m_{a}^{3}\left(n_{a}^{2}+\frac{3m_{a}^2}{m_{d}^{2}}\epsilon_5 n_{d}^{2} \right),\redtwo{\epsilon_3}$)  &  ($\redtwo{\epsilon_4} ,0$)  & ($0,-\redtwo{\epsilon \epsilon_3 \epsilon_4}$)\\
d &  ($\redtwo{\epsilon_5},0$)  &  ($n_{d}^2,\redtwo{\epsilon_{5}}m_{d}^2$)  & ($-\frac{\redtwo{3 \epsilon}}{m_{d}^{2}},-3\frac{m_{a}^{2}}{m_{d}^{2}} m_{a}^{3}$)\\
\hline
\end{tabular}
\caption{\blue{0til-SM models. All of the tori $T^2_i$ are \redtwo{rectangular}. 
The integer parameters denoted by $\epsilon_i$ are $\pm1$. $n_{a}^{2},m_{a}^{3},n_{d}^{2}$ are arbitrary integer numbers and 
$m_{a}^{2},m_{d}^{2}$ are divisors of 3. 
}}
\label{tab:0til-SM}
\end{center}
\end{table}

\begin{table}[h]
\begin{center}
\begin{tabular}{|c||c|c|c|}
\hline
D-brane �& $T_{1}^{2}$ & $T_{2}^{2}$ & $T_{3}^{2}$\\
\hline
\hline
a &  ($\redtwo{1}$,0)  &  ($a/\beta +1,\beta m_{a}^2$)  & ($-\frac{\redtwo{3 \epsilon}}{m_{a}^{2}}, \frac{m_{a}^{3}}{2\beta}$)\\
b &  (0,$ \redtwo{\epsilon_1} $)  &  ($\redtwo{\epsilon_2}/\beta,0$)  & ($0,\redtwo{\epsilon \epsilon_1 \epsilon_2}$)\\
c &  ($-\epsilon_3 \frac{m_a^3}{2\beta} \left( n_a^2 +3 \epsilon_5 n_d^2\frac{m_a^2}{m_d^2} \right),\redtwo{\epsilon_3}$)  &  ($\redtwo{\epsilon_4}/\beta, 0$)  & ($0,\redtwo{-\epsilon \epsilon_3 \epsilon_4}$)\\
d &  ($\redtwo{\epsilon_5} ,0$)  &  ($d/\beta+1,\redtwo{\epsilon_5} \beta m_{d}^{2}$)  & ($-\frac{\redtwo{3 \epsilon}}{m_{d}^{2}},-3\frac{m_a^2}{m_d^2}\frac{m_{a}^{3}}{2\beta}$)\\
\hline
\end{tabular}
\caption{1til-SM models,  $\beta\in \{1,1/2\}$, If $\beta=1$,$T_{3}^{2}$ is the tilted torus and the others are untilted. 
If $\beta =1/2$, $T_{2}^{2}$ is the tilted torus and the others are untilted. 
The integer parameters denoted by $\epsilon_i$ are $\pm1$.
$a,d,m_a^3$ are arbitrary integer numbers and  $m_{a}^{2},m_{d}^{2}$ are divisors of 3.
}
\label{tab:1til-SM}
\end{center}
\end{table}

\begin{table}[h]
\begin{center}
\begin{tabular}{|c||c|c|c|}
\hline
 D-brane  & $T_{1}^{2}$ & $T_{2}^{2}$ & $T_{3}^{2}$\\
\hline
\hline
a &  ($\redtwo{1},0$)  &  ($n_{a}^2,\frac{m_{a}^2}{2}$)  & ($-\frac{3\epsilon}{m_{a}^{2}},m_{a}^{3}$)\\
b &  (0,$\redtwo{\epsilon_1}$)  &  ($\redtwo{2\epsilon_2},0$)  & (0,$\redtwo{\epsilon \epsilon_1 \epsilon_2}$)\\
c &  ($\redtwo{\frac{\epsilon_3 m_{a}^{3}}{2}\left( n_a^2+ \epsilon_5 \frac{m_{a}^{2}}{m_{d}^{2}}n_{d}^{2} \right)},\epsilon_3$)  &  ($\redtwo{2\epsilon_4} ,0$)  & ($0,-\redtwo{\epsilon \epsilon_3 \epsilon_4}$)\\
d &  ($\redtwo{\epsilon_5},0$)  &  ($n_{d}^2,\frac{\redtwo{\epsilon_5} m_{d}^2}{2}$)  & ($-\frac{\redtwo{3\epsilon}}{m_{d}^{2}},-3\frac{\redtwo{m_a^2}}{\redtwo{m_d^2}}m_{a}^{3}$)\\
\hline
\end{tabular}
\end{center}
\caption{2til-SM models. $T_{2,3}^{2}$ are tilted torus and \redtwo{$T_1^2$} is \redtwo{untilted}. 
The  integer parameters denoted by $\epsilon_i$ are $\pm1$.
$n_{a}^{2},n_{d}^{2}$ are arbitrary odd numbers and $m_a^3$ is arbitrary integer number.
$m_{a}^{2},m_{d}^{2}$ are divisors of 3.
}
\label{tab:2til-SM}
\end{table}

The Higgs bosons correspond to the open string in the NS-sector stretching between the D6$_{b}$-brane and the D6$_{c}$-brane. 
These branes are parallel on $T_{2}^{2}$ and $T_{3}^{2}$.
This situation is the same as that in the IMR model \cite{Ibanez:2001nd}.
The Higgs mass is determined by the distance of D6-branes and the intersecting angles.
\red{Note that}
we need fine tuning to get a light Higgs mass.

The D-brane configurations in  Tables \ref{tab:0til-SM}, \ref{tab:1til-SM}, and \ref{tab:2til-SM} do not satisfy the RR tadpole condition yet, 
but this is always possible by adding extra D6-branes which are parallel to the O6-planes.\footnote{These branes can not have couplings with $B_{2}^{k}$ and do not affect massless U(1)s. (See appendix \ref{app:systematic}).}
Since D6$_{a,b,c,d}$-branes and their orientifold mirrors have no intersection points with the O6-plane, 
 there are no intersection points between the extra D-branes and the D6$_{a,b,c,d}$-branes.
Thus, the introduction of these extra D6-branes does not change the chiral spectrum in the visible sector.
In this sense, the extra D6-branes correspond to the completely hidden sector.

These models have characteristic winding numbers.
The D6$_{b}$-brane and the D6$_{c}$-brane are parallel to the O6-plane in $T_{2}^{2}$ and perpendicular to it on $T_3^2$.
The D6$_{a}$-brane and the D6$_{d}$-brane are parallel to the O6-plane in $T_{2}^{1}$.
The charge of $U(1)_{a}$ is 3 times  the baryon number and the $U(1)_{d}$ charge is the lepton number.
The \redtwo{intersection} numbers between the D6$_{a,c}$-brane and the D6$_{b,c}$-brane in $T_{2,3}^{2}$ are the same.
Thus, the flavor structure of the quarks and leptons are exactly the same at perturbative level.
(See for discrete flavor symmetries \cite{Abe:2009vi,BerasaluceGonzalez:2012vb}.)\footnote{
Similarly flavor symmetries are obtained in heterotic orbifold models \cite{Kobayashi:2004ya}. 
 See also \cite{Higaki:2005ie} .}
However, if we take  non-perturbative effects into account, these structures must be broken and, 
for example, right-handed Majorana neutrino masses might be generated \cite{Blumenhagen:2006xt,Ibanez:2006da,Hamada:2014hpa}.
At any rate, the study of the flavor sector is beyond our scope at this time.

\section{Gauge couplings}
\label{sec:gauge}

\subsection{Model constraints}
\label{subsec:constraint}

We have found three classes of SM-like models in section 3.
In these models, the gauge symmetry is exactly the same as that of the SM up to the hidden sector.
Now, let us study the gauge sector quantitatively.
That is, we study the question whether it is possible to make 
all gauge couplings consistent with their experimental values.
At first sight, it appears possible because there are a lot of parameters in these classes of models.
For example, all classes of models have torus moduli and more than three integer winding numbers as free parameters.\footnote{Precisely speaking, we need to consider the stabilization of the moduli. However, this issue is beyond the scope of this paper and we treat the moduli as free parameters.}
However, it becomes more complicated when we take into account other constraints.
One constraint is to avoid the tachyonic configurations and the other is a constraint on the string coupling.

The R-sector of the open string stretching between D-branes has a chiral fermionic zero-mode, while the corresponding NS-sector 
has the light scalar spectrum of Table \ref{tab:NS_light_spectrum}.
These NS-sector modes are the superpartners of the chiral fermions and some of them could be tachyonic in non-supersymmetric models.
If a configuration has tachyons, it is unstable and 
decays to another configuration quickly.
We must tune parameters to avoid such tachyons.
This condition constrains the parameters significantly.
In $Sp(2)$ models, there are six chiral fermion modes and each of them has superpartners at \redtwo{intersection} points.
To make these scalars massive or massless, the models must satisfy 24 inequalities.

The other constraint is the perturbativity of the theory.
The tree level gauge coupling $\alpha_k=g_{k}^2/4\pi$ at the string scale is given by \cite{Klebanov:2003my,Blumenhagen:2003jy},
\begin{equation}
\frac{1}{\alpha_{k}}=\frac{M_{s}^{3}V_{k}}{(2\pi)^{3}g_{s}\kappa_{k}},
\label{eq:gauge_coupling}
\end{equation}
where $V_{k}$ denotes the D$6_k$-brane's 3-cycle volume in the compact space, $M_{s}$ is the string scale and $g_{s}$ is the string coupling.
$\kappa_{k}$ is obtained as 
$\kappa_{k}=1$ for $U(N_{k})$ and $\kappa_{k}=2$ for $Sp(2N_{k})/SO(2N_{k})$.
In this way, we can calculate all the gauge couplings, $\alpha_{a,b,c,d}$.
For $U(1)_Y$, we must normalize the gauge field and $\alpha_{Y}$ is written by,
\begin{equation}
\frac{1}{\alpha_{Y}}=\frac{1}{6} \frac{1}{\alpha_{a}}+\frac{1}{2} \frac{1}{\alpha_{c}}+\frac{1}{\blue{2}} \frac{1}{\alpha_{d}}.
\label{eq:gauge-Y}
\end{equation}

On the other hand, \red{by performing dimensional reduction of the type IIA supergravity action,} one can write the Planck mass $M_{p}$ using string parameters as,
\begin{equation}
M_{p}^{2}=\frac{8M_{s}^{8} V_{6}}{(2\pi)^{6} g_{s}^{2}},
\label{eq:planck_scale}
\end{equation}
where $V_{6}$ is the volume of the compact space.
{}From (\ref{eq:gauge_coupling}), (\ref{eq:planck_scale}), we can write the string coupling  in terms of gauge couplings, 
\begin{equation}
g_{s} = \frac{\alpha_{k}^{4}}{8^{3/2} (2\pi)^{3} \kappa_{k}^{4}} \left( \frac{V_{k}^{2}}{V_{6}} \right)^{2} V_{6}^{1/2} M_{p}^{3}.
\label{eq:gst}
\end{equation}
We have  concentrated on perturbative vacua and their effective theories, but when \red{$g_{s} \blue{>} \mathcal{O}(1)$}, perturbative theory is broken down and our models no longer make sense.
To get sufficiently small $g_{s}$, there are constraints on parameters.

It is natural to assume $V_{6} \sim 1/M_{s}^{6}$.
The $\alpha_{k}$ in Eq.~(\ref{eq:gst}) is the gauge coupling at the string scale, so we evaluate  
\begin{equation}
g_{s} \sim 2\times10^{-4}\frac{\alpha_{k}(M_{s})^{4}}{\kappa_{k}^{4}} \left( \frac{V_{k}^{2}}{V_{6}} \right)^{2}  \left( \frac{M_{p}}{M_{s}}\right)^{3}.
\label{eq:gst1.1}
\end{equation}
Naively, if $M_{s}$ is very small, $g_{s}$ is very large and perturbativity of the theory is violated.

Using the renormalization group equations and the experimental values of $\alpha_k(M_Z)$, 
we can evaluate $\alpha_{k}(M_{s})$ in Eq.~(\ref{eq:gst1.1}).
The models obtained in the previous section have almost the same field contents as those of the SM, but 
include gauginos and adjoint scalars in the gauge sector.
We assume that such gauginos and adjoint scalars gain masses around $M_s$ and neglect their threshold corrections.\footnote{For more precise comments, see Appendix B.}
Hence, we can evaluate $\alpha_{k}(M_{s})$ by using beta-functions of the SM.
{}We find $\alpha_{3,2}(M_{s}) > 1/50$ for $M_s \leq 10^{18}$ GeV.
Then, $V_{a,b}/(V_{6})^{\frac{1}{2}}$ must be small to get \red{sufficiently} small $g_{s}$.
This means that the direction which is perpendicular to the a,b-brane is large and $V_{a,b}/V_{6}$ is suppressed.
However, in our models, we have $I_{ab}\neq 0$ and there is no direction which is perpendicular to a-brane and b-brane at the same time.
Hence, generally we get $V_{a}V_{b}/V_{6} > 1$.
When $V_{a}V_{b}/V_{6} > 1$ and $\alpha_{3},\alpha_{2} > 1/50$, we obtain 
\begin{equation}
\begin{split}
g_{s} &\sim 2\times10^{-4} \alpha_{3}(M_{s})^{2}\alpha_{2}(M_{s})^{2} \left( \frac{V_{a}V_{b}}{V_{6}} \right)^{2}  \left( \frac{M_{p}}{M_{s}}\right)^{3},\\
&\gtrsim 10^{-12} \left( \frac{M_{p}}{M_{s}}\right)^{3}.
\end{split}
\label{eq:string_scale}
\end{equation}
This requires $M_{s} \gtrsim10^{15}$GeV.
When there is a large hierarchy between $V_6$ and $1/M_s^6$, 
this estimation would change.
For $V_6M_s^6= \gamma$, we have the constraint  $M_{s} \gtrsim \gamma^{1/6} 10^{15}$GeV.
For example, we find $M_s \gtrsim10^{16}$GeV for $\gamma = {\cal O} (10^{6})$ and 
$M_s \gtrsim10^{14}$GeV for $\gamma = {\cal O} (10^{-6})$.
We should comment on the effect of the gauginos and adjoint scalars on the previous argument.
We have assumed that all of the gauginos and adjoint scalars have masses around $M_s$.
If they are lighter, $\alpha_{3}$ and $\alpha_{2}$ become larger because they give positive contributions to beta-functions.
Therefore, the lighter gauginos and adjoint scalars strengthen the constraint.

As mentioned above,  the string scale is constrained.
On the other hand, winding numbers and moduli are also  constrained.
As a concrete example, we study the 0til-SM models.
In this class of models, the ratio of tree level gauge couplings is given by,
\begin{equation}
\begin{split}
\frac{1}{\alpha_{3}} : \frac{1}{\alpha_{2}}
&=
{\rm Re}~\tau_{1}\sqrt{(n_a^2 {\rm Re}~\tau_{2})^2+(m_{a}^{2})^{2} }\sqrt{\left(\frac{3}{m_{a}^{2}} {\rm Re}~\tau_{3}\right)^2+(m_{a}^{3})^2}
: {\rm Re}~\tau_{2},\\
&=
{\rm Re}~\tau_{1}\sqrt{(n_a^2 )^2+(m_{a}^{2}/{\rm Re}~\tau_2)^{2} }\sqrt{\left(\frac{3}{m_{a}^{2}} 
{\rm Re}~\tau_{3}\right)^2+(m_{a}^{3})^2}
: 1,
\end{split}
\end{equation}
where $\tau_{i}$ is the $T^2_i$ torus modulus.
The renormalization group flows from the experimental values show that 
 $\alpha_{2}(\mu)$ is similar to $\alpha_{3}(\mu)$ unless the running scale $\mu$ is very low.
To realize $\alpha_{2}(M_s) \sim \alpha_{3}(M_s)$, it is required that $|\tau_{1}|$ is less than ${\cal O}(1)$.
In this way, the \red{winding numbers and the value of the moduli}
are constrained.

In supersymmetric models, stringy one-loop threshold corrections have been calculated \cite{Lust:2003ky,Akerblom:2007uc,Gmeiner:2009fb}, 
and they can be sizable\footnote{See e.g. \cite{Honecker:2012qr}.} for large values of moduli.
On the other hand, threshold corrections have not been calculated in non-supersymmetric models.
We assume that such threshold corrections are sub-dominant compared with the tree-level values, $\alpha_a(M_s)$.
Otherwise, higher order corrections would also be large and perturbativity would be violated.
Thus, the above estimations are valid under the assumption that stringy threshold corrections are 
sufficiently smaller than the tree-level values.
In the next subsection, we study the gauge couplings numerically while neglecting stringy threshold corrections.\footnote{
See Appendix B for estimation of threshold corrections in a model.}

\subsection{Numerical analysis}
We plot the gauge coupling ratios of our models in Figures \ref{fig:gc16GeV}, \ref{fig:gc15GeV} and \ref{fig:gc14GeV} for $M_s = 10^{16},$ $10^{15}$ and $10^{14}$ GeV, respectively.
For comparison, we also show the gauge coupling ratios of the IMR model in these figures.
The blue data points correspond to the gauge coupling ratios, which are calculated 
by Eqs.~(\ref{eq:gauge_coupling}) and  (\ref{eq:gauge-Y}) for the parameters to satisfy $g_s <1$ \red{assuming} $V_6 = 1/\red{M_s^6}$ and 
to avoid tachyonic modes.
Moduli should be stabilized, but we used them as free parameters.
We vary winding numbers from 1 to 100 and torus moduli from $10^{-2}$ to $10^2$.
There are two types of modes. One is localized at intersection points on all of the three
$T^2$, and the other is stretching between parallel D-branes on one or two of the three
$T^2$.
For the first type of modes, we vary the parameters of our models, the moduli and
the winding numbers, such that non of them are tachyonic.
For the second mode, we make them massless or massive by tuning open string moduli.
Note that the ratios \red{$\alpha_k/\alpha_l$} given by Eqs.~(\ref{eq:gauge_coupling}) and  (\ref{eq:gauge-Y}) are independent of 
$M_s$.
Thus, if we do not impose other constraints, the same blue data points (gauge coupling ratios) would appear for $M_s = 10^{\blue{14}}, 10^{\blue{15}}$ and $10^{\blue{16}}$ GeV.
However, the constraint $g_s <1$ depends on $M_s$.
The constraint becomes severe for a lower $M_s$.
That is, the difference between these figures only comes from the perturbativity condition.
Obviously, it is more constrained in Figures \ref{fig:gc15GeV} and \ref{fig:gc14GeV} 
and the number of blue data points is less than that in Figure \ref{fig:gc16GeV}.
The red data points correspond to the \red{$\overline{\text{MS}}$} renormalized gauge coupling ratios of the SM \red{computed by using} the experimental values, 
i.e. $\alpha_3(\mu)/\alpha_Y(\mu)$ and $\alpha_2(\mu)/\alpha_Y(\mu)$.
{}From top to bottom, the data points represent $\mu = 10^3, 10^4, \cdots 10^{19}$ GeV.
The model can fit the gauge couplings if the blue data points overlap with the red data points corresponding to $\mu = M_s$, $\mu= 10^{\blue{16}}$ GeV in  
Figure \ref{fig:gc16GeV},   $\mu= 10^{\blue{15}}$ GeV  in Figure \ref{fig:gc15GeV} and $\mu= 10^{\blue{14}}$ GeV in Figure \ref{fig:gc14GeV}.

\setcounter{figure}{-4}
\begin{figure}[ht]
\captionsetup{labelformat=empty,labelsep=none}
\begin{tabular}{cc}
\begin{minipage}{0.55\hsize}
\begin{center}
\epsfig{file=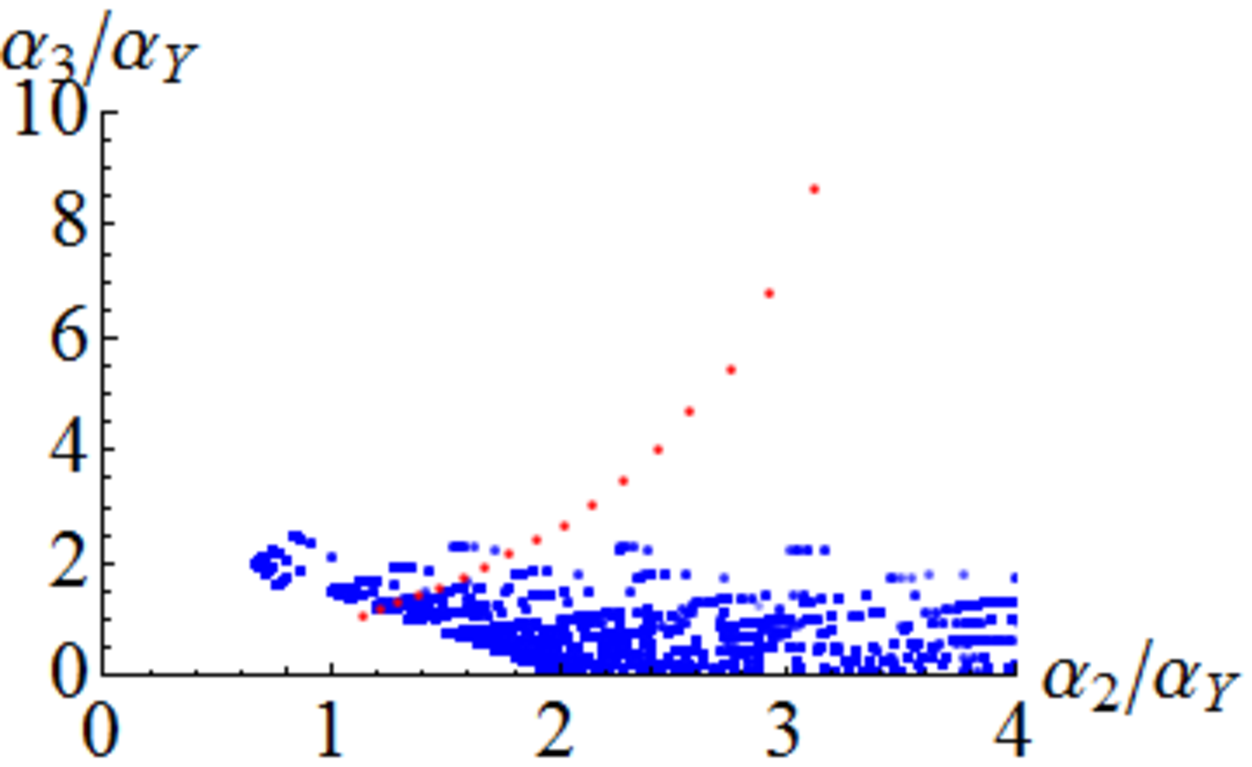,scale=0.7}
\caption{0til-SM}
\end{center}
\end{minipage}
\begin{minipage}{0.55\hsize}
\begin{center}
\epsfig{file=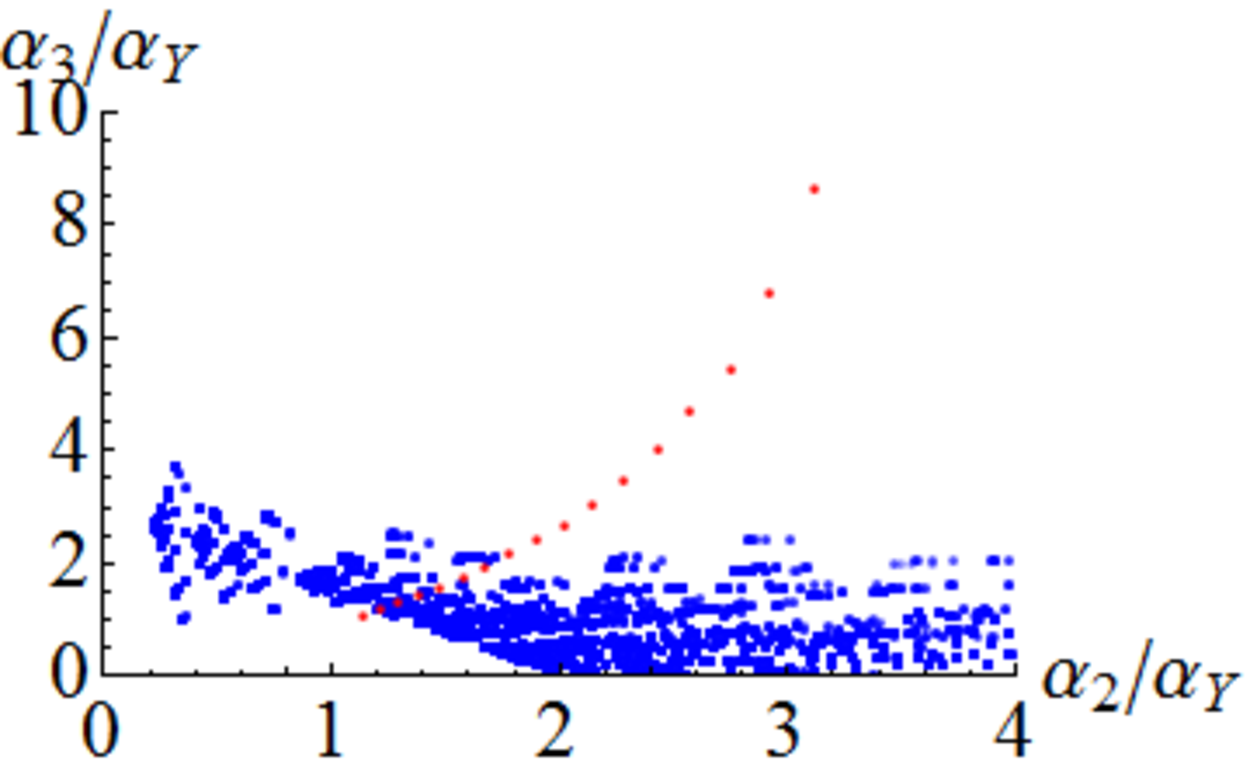,scale=0.7}
\caption{1til-SM}
\end{center}
\end{minipage}
\\
\begin{minipage}{0.55\hsize}
\begin{center}
\epsfig{file=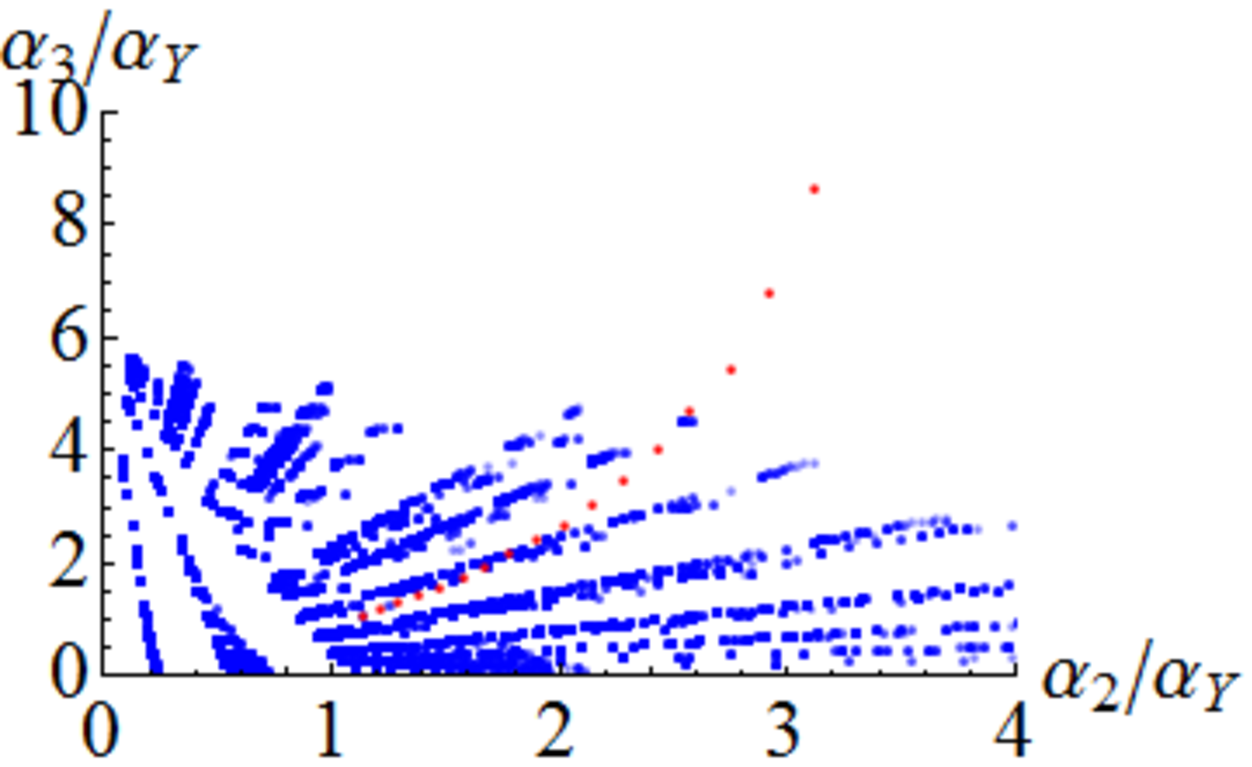,scale=0.7}
\caption{2til-SM}
\end{center}
\end{minipage}
\begin{minipage}{0.55\hsize}
\begin{center}
\epsfig{file=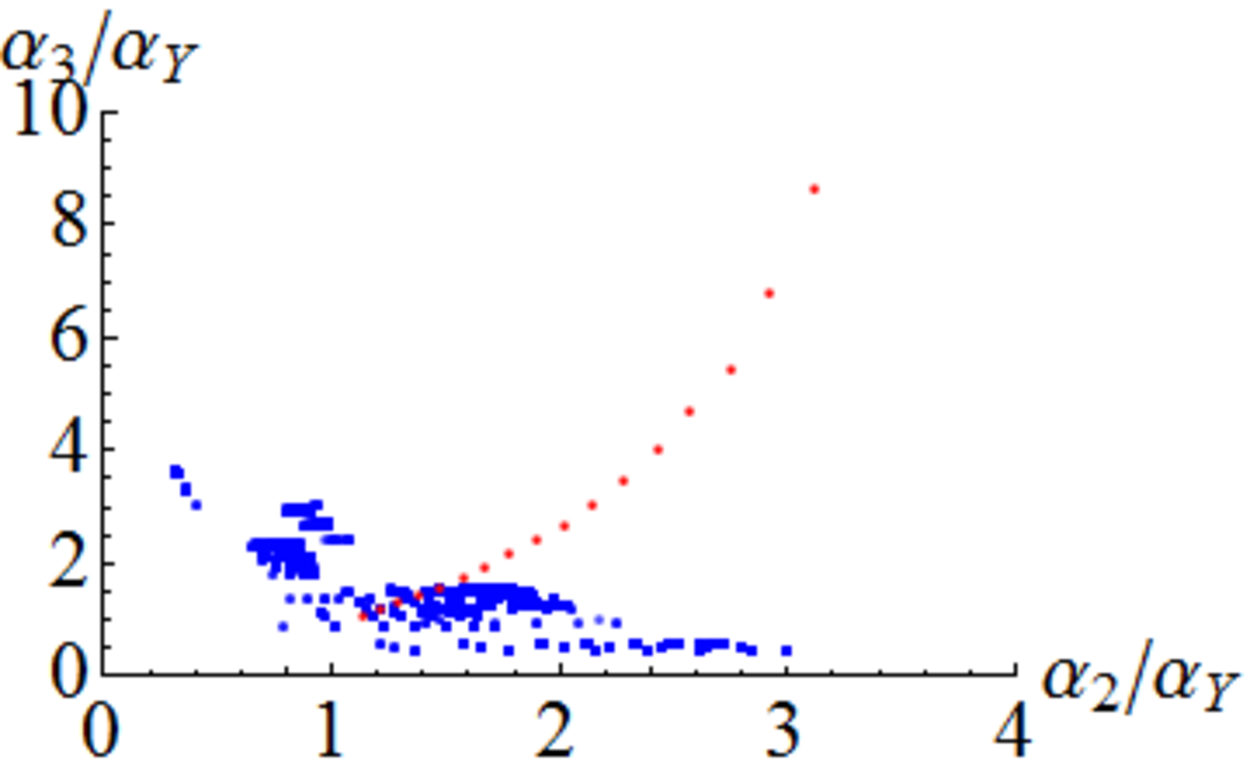,scale=0.7}
\caption{The IMR model}
\end{center}
\end{minipage}
\end{tabular}
\captionsetup{labelformat=default,labelsep=colon}
\caption{Distributions of the ratio of gauge couplings.
The blue data points are the gauge coupling ratios of $Sp(2)$ models and the model in \cite{Ibanez:2001nd} and the red data points are renormalized gauge couplings of the SM.
The red data point to the upper right is the renormalized gauge coupling at $10^{3}$ GeV and lower left data points is at $10^{19}$ GeV.
The winding numbers range from 1 to 100 and the torus moduli from $10^{-2}$ to $10^2$.
We set $M_{s}$ to \blue{$10^{16}$GeV} and non-perturbative configurations are eliminated.
Tachyon configurations are eliminated, too.
}
\label{fig:gc16GeV}
\end{figure}

\setcounter{figure}{-3}
\begin{figure}[ht]
\captionsetup{labelformat=empty,labelsep=none}
\begin{tabular}{cc}
\begin{minipage}{0.55\hsize}
\begin{center}
\epsfig{file=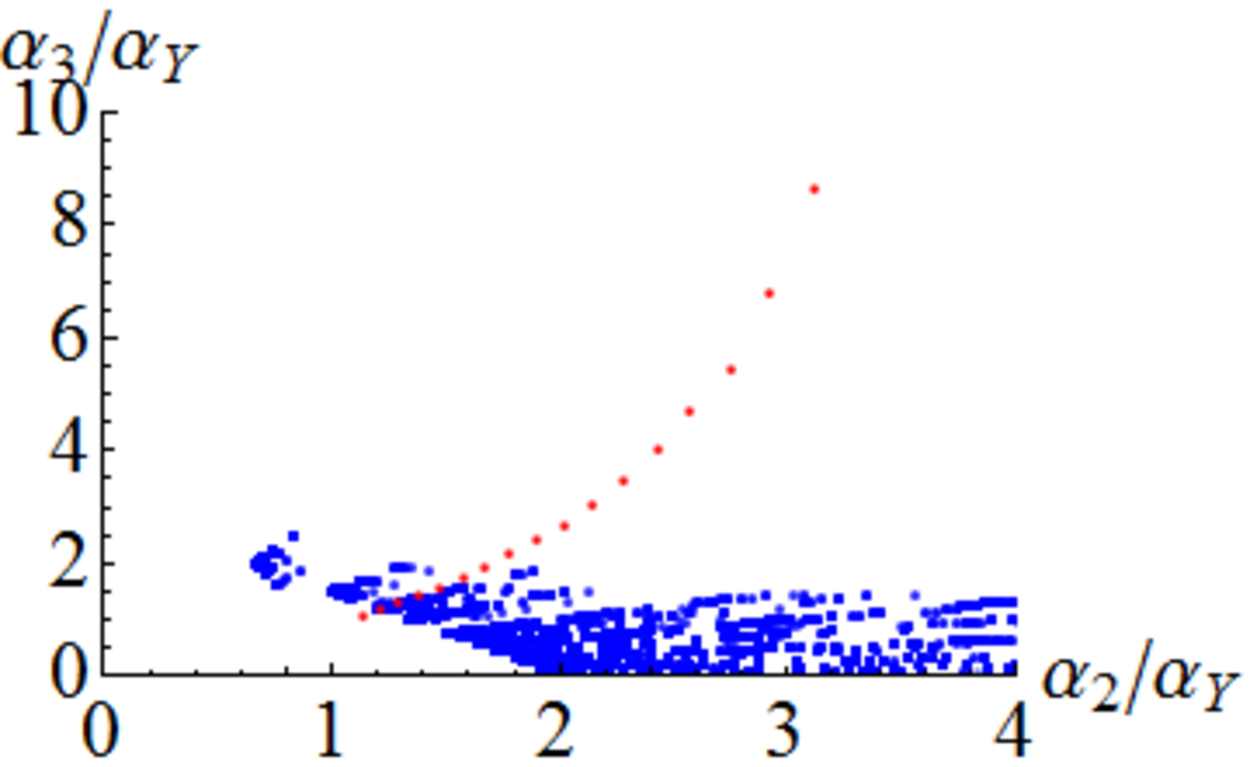,scale=0.7}
\caption{0til-SM}
\end{center}
\end{minipage}
\begin{minipage}{0.55\hsize}
\begin{center}
\epsfig{file=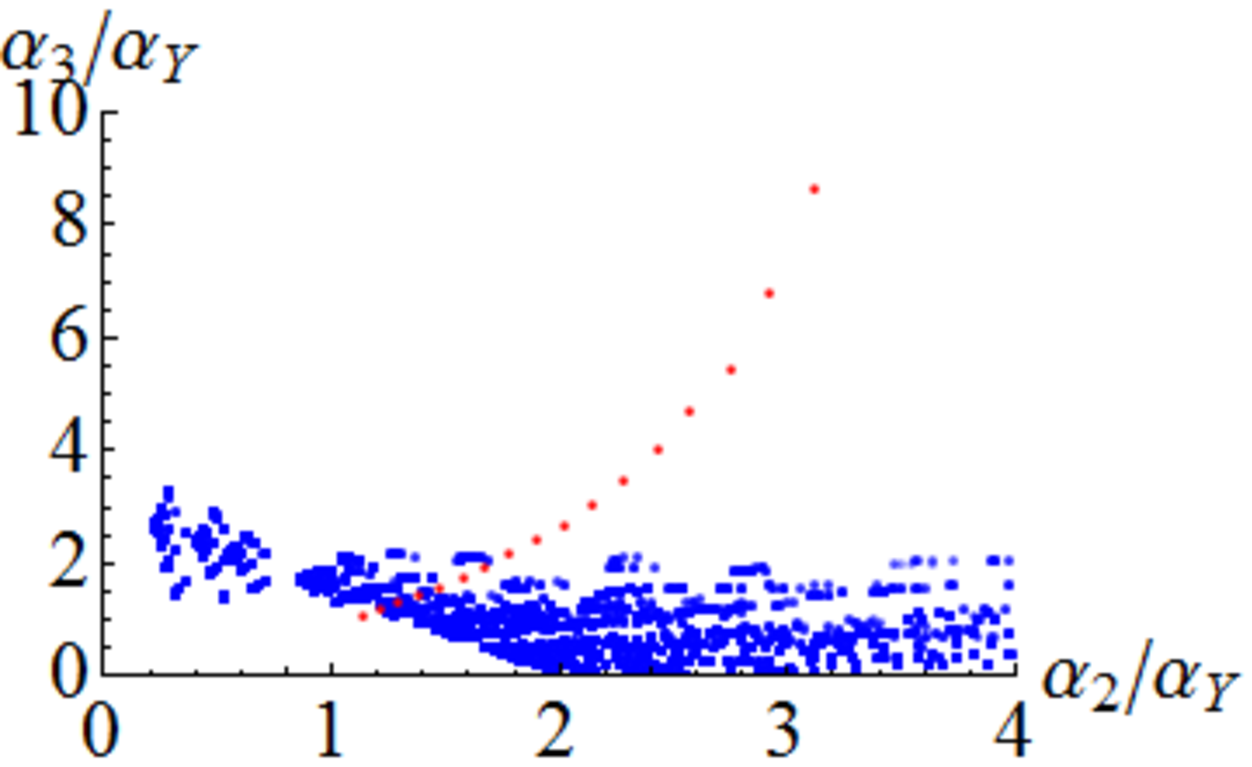,scale=0.7}
\caption{1til-SM}
\end{center}
\end{minipage}
\\
\begin{minipage}{0.55\hsize}
\begin{center}
\epsfig{file=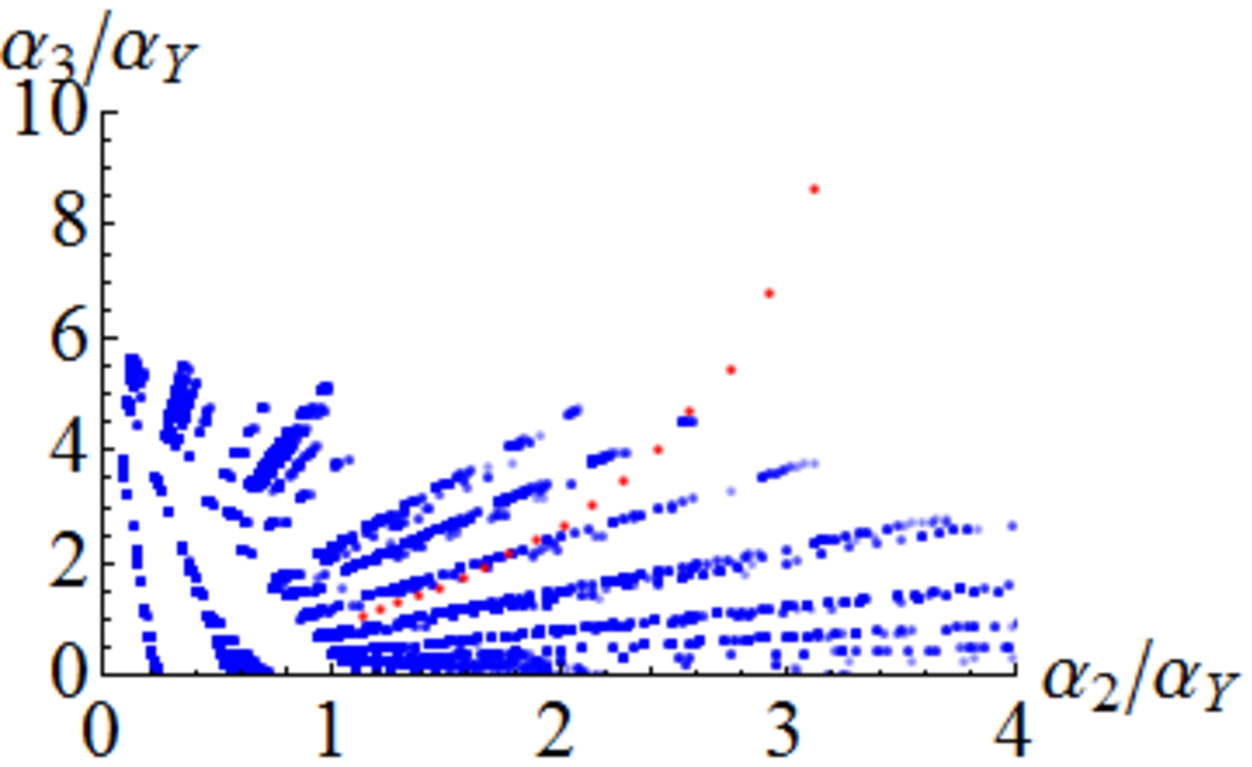,scale=0.7}
\caption{2til-SM}
\end{center}
\end{minipage}
\begin{minipage}{0.55\hsize}
\begin{center}
\epsfig{file=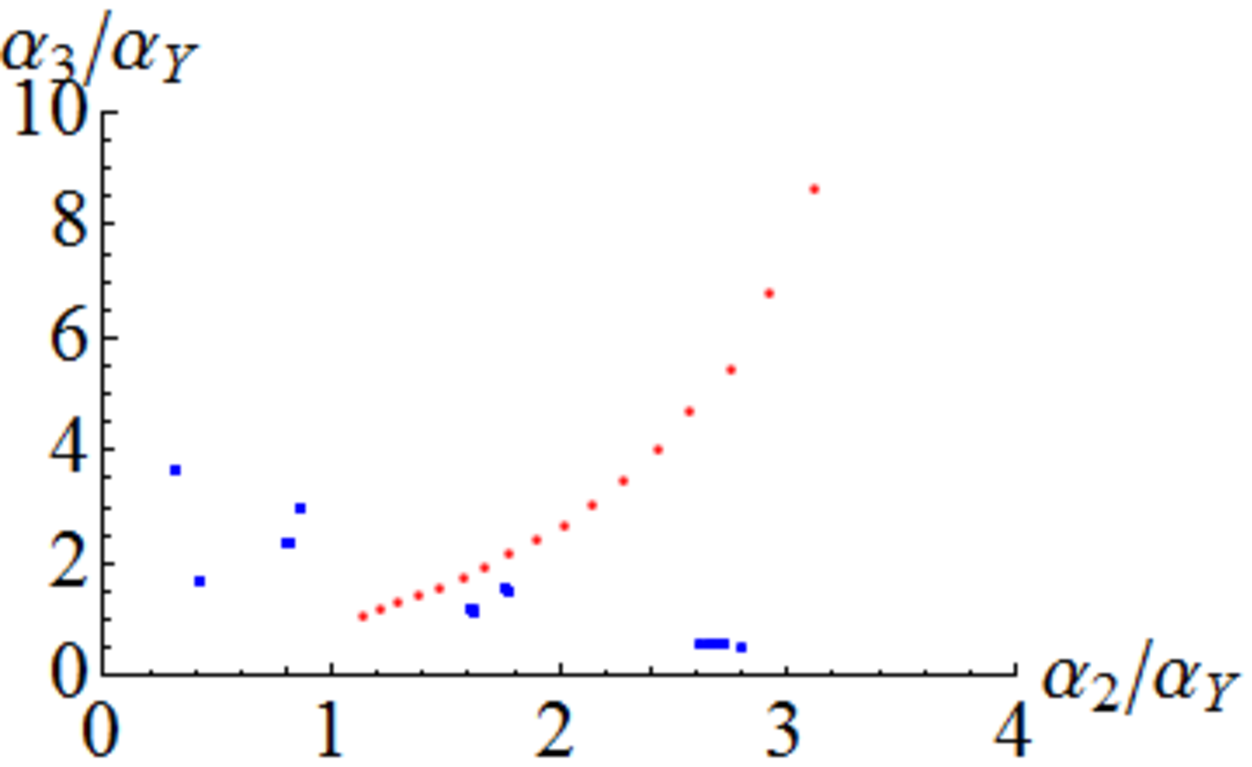,scale=0.7}
\caption{The IMR model}
\end{center}
\end{minipage}
\end{tabular}
\captionsetup{labelformat=default,labelsep=colon}
\caption{Distributions of the ratio of gauge couplings.
The winding numbers and the torus moduli are not changed from Figure \ref{fig:gc16GeV}.
In this figure, we set $M_{s}$ to \blue{$10^{15}$ GeV}.
}
\label{fig:gc15GeV}
\end{figure}

\setcounter{figure}{-2}
\begin{figure}[ht]
\captionsetup{labelformat=empty,labelsep=none}
\begin{tabular}{cc}
\begin{minipage}{0.55\hsize}
\begin{center}
\epsfig{file=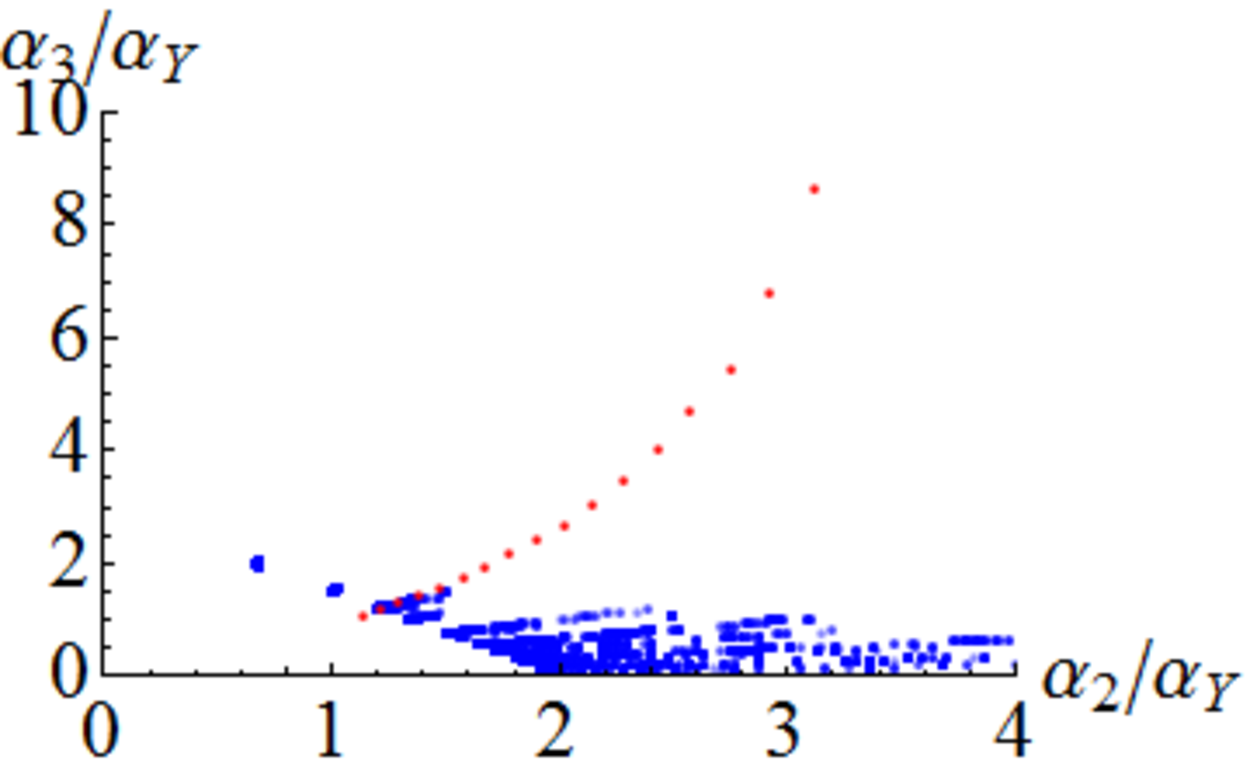,scale=0.7}
\caption{0til-SM}
\end{center}
\end{minipage}
\begin{minipage}{0.55\hsize}
\begin{center}
\epsfig{file=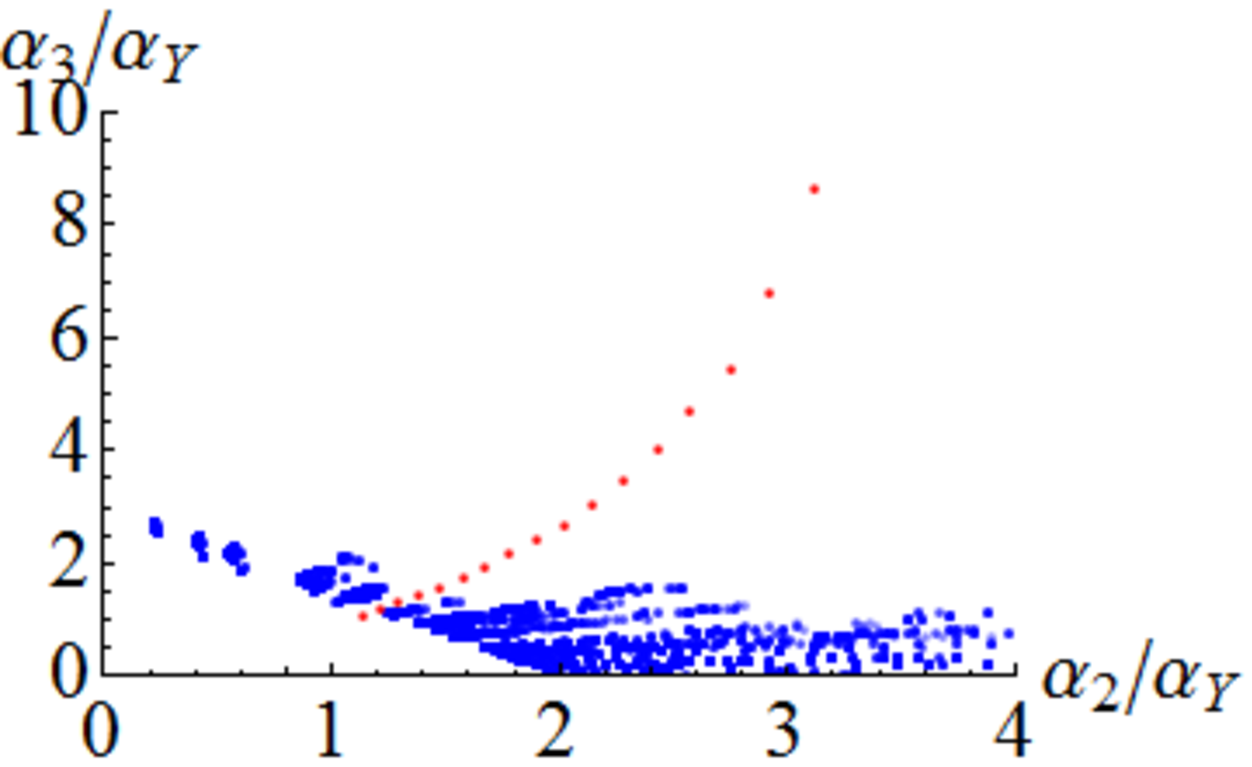,scale=0.7}
\caption{1til-SM}
\end{center}
\end{minipage}
\\
\begin{minipage}{0.55\hsize}
\begin{center}
\epsfig{file=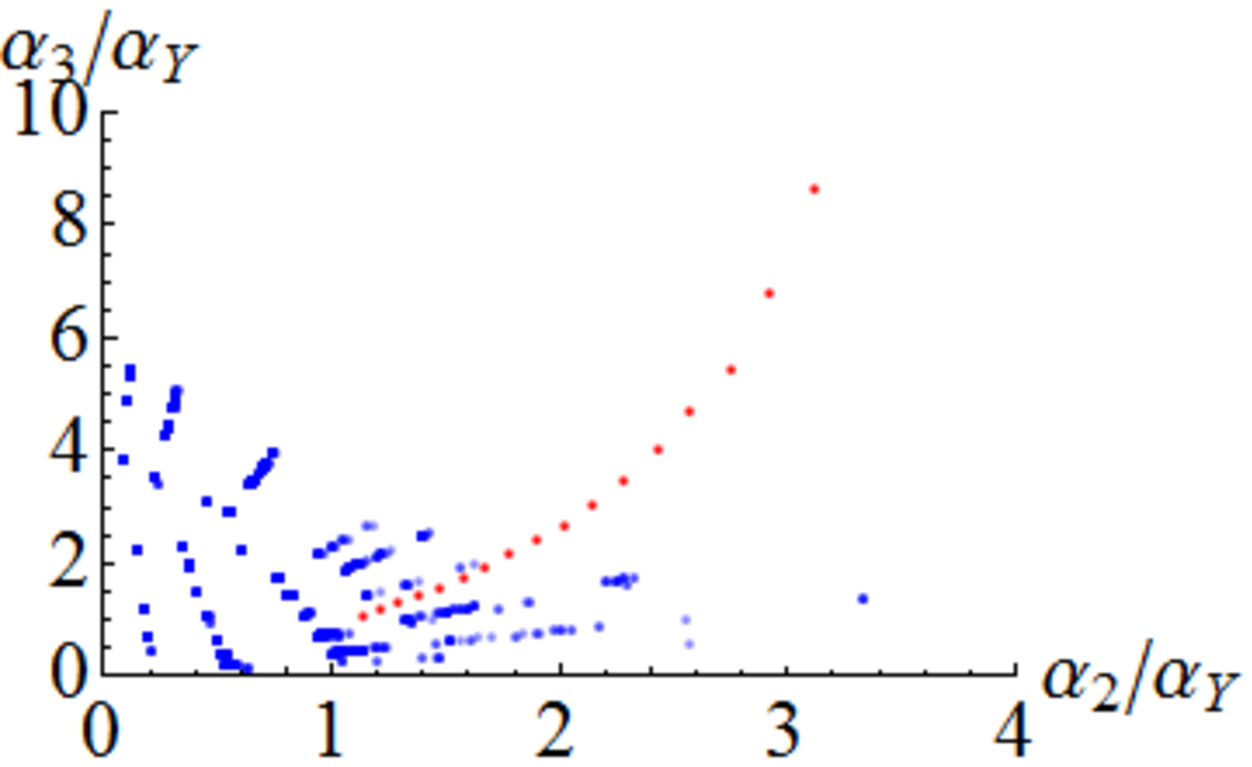,scale=0.7}
\caption{2til-SM}
\end{center}
\end{minipage}
\begin{minipage}{0.55\hsize}
\begin{center}
\epsfig{file=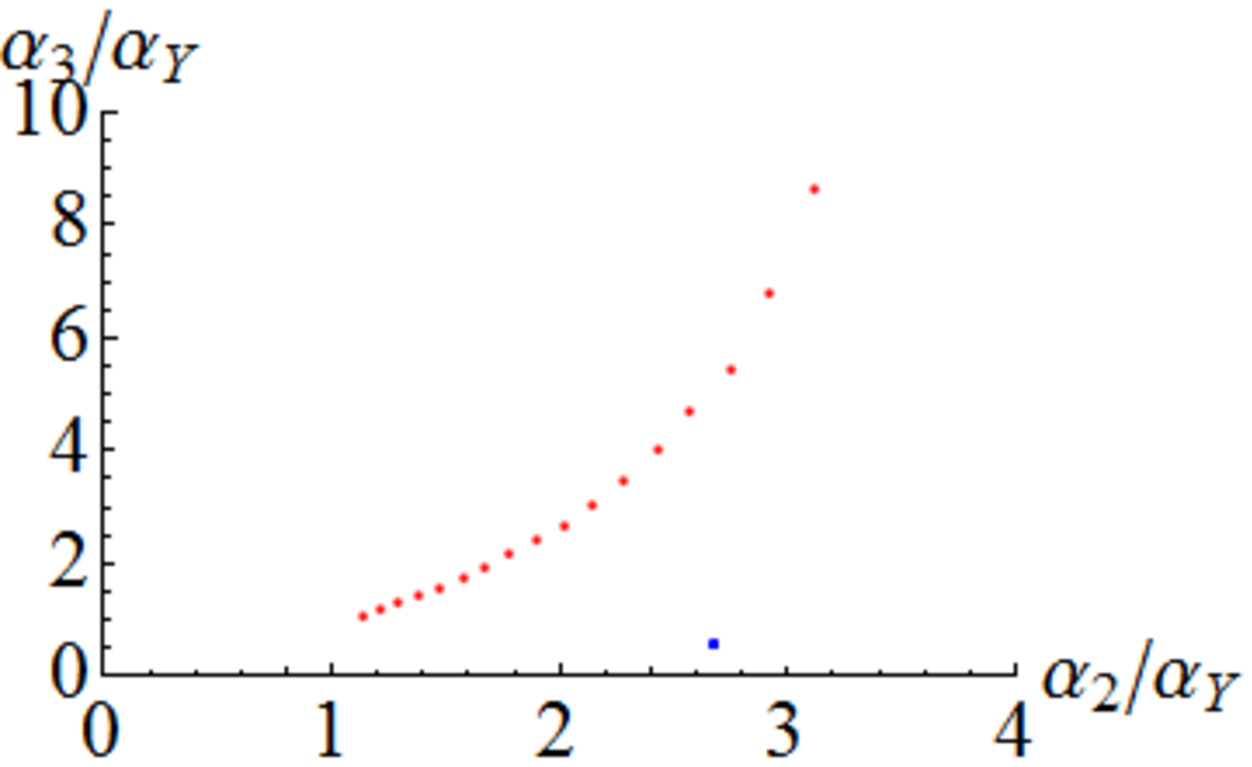,scale=0.7}
\caption{The IMR model}
\end{center}
\end{minipage}
\end{tabular}
\captionsetup{labelformat=default,labelsep=colon}
\caption{Distributions of the ratio of gauge couplings.
The winding numbers and the torus moduli are not changed from Figure \ref{fig:gc16GeV}.
In this figure, we set $M_{s}$ to $10^{14}$GeV.
}
\label{fig:gc14GeV}
\end{figure}

There are some characteristic \red{features}
in these figures.
In all models, the ratio of the gauge couplings $\alpha_{3}/\alpha_{Y}$ is less than \blue{6}.
This is because $U(1)_{Y}$ is a linear combination of $U(1)_{a,c,d}$s and $\alpha_{Y}$ is function of $\alpha_{3}$.
It leads to an upper bound on $\alpha_{3}/\alpha_{Y}$.
$Sp(2)$ models tend to have larger $\alpha_{2}$ than $U(2)$ model.
This is because the b-brane must be parallel or perpendicular to the O6-plane in $Sp(2)$ and its volume can not be so large.
The $Sp(2)$ models have a larger allowed region than the IMR model. This is because the $Sp(2)$ models have more parameters than the IMR model.

Figure \ref{fig:gc16GeV} shows that we can tune parameters to fit gauge couplings in  all models to the experimental values if $M_{s}$ is greater than  $10^{16}$GeV.
For $M_{s}=10^{15}$GeV, we can realize the gauge couplings in $Sp(2)$ models. For the IMR model, there are no blue data points overlapping red data points, but we would find suitable parameters explaining the experimental values by a more dense parameter search. 
For $M_{s}=10^{14}$ GeV, we can explain experimental values in 2til-SM models and it would be possible in the other $Sp(2)$ models. 
We checked that blue data points disappear in this region for $M_{s}=10^{13}$GeV and we can not tune parameters to fit the gauge couplings for weak $g_s$ in any of these models.
The critical string scale is $10^{14-15}$GeV.
These results are consistent with Eq.~\eqref{eq:string_scale}.

In our analysis, we assumed $V_6M_s^6= 1$.
Similarly, we can analyze gauge couplings for other values of $V_6M_s^6= \gamma$.
Unless there is a large hierarchy between them, we obtain almost the same results.
Furthermore, \blue{even} when $\gamma$ is very small or large, we would have the lower bound on $M_s$.
In some cases, the one-loop threshold corrections would become significant \cite{Lust:2003ky}.

\subsection{Explicit example}
\label{subsec:explicit}

\red{In this subsection}, we give an explicit example of one of the models.
As shown in Figure \ref{fig:gc16GeV}, there are a lot of winding numbers and moduli which realize the renormalized SM gauge couplings at the string scale. 
Table \ref{tab:example} shows one example.

\begin{table}
\begin{center}
\begin{tabular}{|c||c|c|c|}
\hline
  D-brane & $T_{1}^{2}~(1/{\rm Re}\tau_1=10^{2/3})$ & $T_{2}^{2}~(1/{\rm Re}\tau_2=10^{14/9})$ & $T_{3}^{2}~(1/{\rm Re}\tau_3=10^{2/3})$\\
\hline
\hline
a &  (1,0)  &  ($3,1/2$)  & ($-3,1/2$)\\
b &  (0,$1$)  &  (2,0)  & (0,1)\\
c &  ($4,1$)  &  ($2 ,0$)  & ($0,-1$)\\
d &  ($1,0$)  &  ($13,3/2$)  & ($-1,-1/2$)\\
\hline
\end{tabular}
\caption{The explicit example of winding numbers and moduli realizing the SM gauge coupling ratio in 2til-SM model.}
\label{tab:example}
\end{center}
\end{table}

In this model, the string scale is set to be $10^{18}$ GeV and the ratios of the gauge couplings in the model are 
given as,  
\begin{equation}
\begin{split}
\alpha_{3}/\alpha_{Y}=1.2,\\
\alpha_{2}/\alpha_{Y}=1.2.
\end{split}
\end{equation}
{}From the experimental values, the ratios of renormalized gauge couplings at $10^{18}$ GeV are,
\begin{equation}
\begin{split}
\alpha_{3,{\rm ren}}/\alpha_{Y,{\rm ren}}=1.2,\\
\alpha_{2,{\rm ren}}/\alpha_{Y,{\rm ren}}=1.2.
\end{split}
\end{equation}
To get the realistic gauge couplings, the string coupling should be $5\times 10^{-3}$, which means that the theory is weakly coupled.

\section{Conclusion and discussion}

We have studied SM-like intersecting D-brane models.
We have constructed and classified the simplest class of models 
using $Sp(2)$ which realizes the SM gauge symmetry and chiral spectrum 
including three right-handed neutrinos as open string zero modes.
These models are very simple and attractive.
They have only four stacks of D-branes.
The three generations of leptons and quarks are just realized by intersection numbers of D-branes, 
and each generation originates from the same type of intersection point.
This is different from the IMR model, where one quark doublet generation originates from 
the intersection point between the $D6_a$-brane and the $D6_b$-brane, while the other two generations originate from 
 the intersection point between the $D6_a$-brane and the $D6_{b^*}$-brane. 
Thus, our models have very large flavor symmetry.
Its proper breaking might be helpful to realize the flavor structure found in nature.

We have studied the gauge coupling constants of our models.
At first sight, it seems always possible to fit the gauge couplings to the experimental values in most of models, 
because there are numerous free parameters.
However, it is non-trivial to reproduce the SM gauge couplings because two conditions, the absence of tachyons and perturbativity, put strong constraints on the model parameters. 
Our calculation \red{has shown} that the string scale must be greater than \blue{ 10$^{14-15}$GeV}  to get realistic gauge couplings 
when there is no large hierarchy between $V_6$ and $M_s$.
Low energy strings are disfavored in these models.
This tendency may not be  model-dependent.
One reason is that $\alpha_Y$ must depend on $\alpha_3$ and $\alpha_{3}/\alpha_{Y}$ has some limits in intersecting D-brane models.
When we try to reconstruct the SM, the values of gauge coupling constants have similar values.

In order to fit the gauge couplings to the experimental values, 
we have used moduli parameters as free parameters.
However, moduli should be stabilized and their stabilized values are 
important to realize the gauge couplings.
All of our models include a hidden sector.
Some dynamics in the hidden sector are expected to play a role in moduli stabilization.
Also, the hidden sector may include dark matter.
These topics are quite interesting, but beyond the scope of the work presented here.

\section*{Acknowledgement}

The authors would like to thank Jahn Alexander for his kind comments and advice.
 The work of Y. H. is supported in part by the 
Grant-in-Aid for  Japan Society for the Promotion of Science (JSPS)
Fellows No.25$\cdot$1107. 
The work of T.K. is supported in part by the Grants-in-Aid for  Scientific
 No. 25400252 from the Ministry of Education, Culture, Sports, Science and 
Technology of Japan. 

\appendix 

\section{Systematic analysis of D-brane configurations}
\label{app:systematic}

We study systematically all the possible D-brane configurations 
of four stacks of $D6$-branes, $D6_{a,b,c,d}$ leading 
to the gauge group $SU(3)\times Sp(2) \times U(1)_Y \times G_{hidden}$ and 
the following intersecting numbers:
\begin{equation} I_{ab}=n;\ I_{ac} = -n;\ I_{ac^*}=-n;\ I_{ad}=0;\ I_{ad^*} = 0,\nonumber \end{equation}
\begin{equation} I_{bc}=0;\ I_{db} = n;\ I_{dc}= -n;\ I_{dc^*} = -n,\nonumber \end{equation}
\begin{equation} I_{aa^*}=0;\ I_{cc^*} = 0;\ I_{dd^*}=0,\label{eq:int_n}\end{equation}
where $n$ is the generation number and where we are especially interested in the n=3 case, for obvious reasons.

Since $I_{aa*}=0$ and D6$_{a}$-branes are parallel with the O-plane in one brane to avoid extra zero modes, we can write,
\begin{description}
\item[]$(n_{a}^{1},m_{a}^{1})=(n_{a}^1,0)$,
\end{description}
without loss of generality.
Because $I_{ab}=n$ and the D6$_{b}$-brane is parallel or perpendicular to the O-plane, all the possible D6$_b$-brane configurations are classified as follows,
\begin{description}
\item[(1)]$(n_{b}^{1},m_{b}^{1})=(0,m_{b}^{1}), (n_{b}^{2},m_{b}^{2})=(n_{b}^{2},0), (n_{b}^{3},m_{b}^{3})=(n_{b}^{3}, 0)$,
\item[(2)]$(n_{b}^{1},m_{b}^{1})=(0,m_{b}^{1}), (n_{b}^{2},m_{b}^{2})=(0,m_{b}^{2}), (n_{b}^{3},m_{b}^{3})=(0,m_{b}^{3})$,
\item[(3)]$(n_{b}^{1},m_{b}^{1})=(0,m_{b}^{1}), (n_{b}^{2},m_{b}^{2})=(n_{b}^{2},0), (n_{b}^{3},m_{b}^{3})=(0,m_{b}^{3})$,
\end{description}
\blue{where $n_{b}^{i}$s and $m_{b}^{i}$s are integers.} 
\blue{Since $I_{ab}$ is proportional to $n_{a}^{1}\cdot m_{b}^{1}$ and $|n_{a}^{1}|$ is even with ${\rm Im}\tau_{1}=1/2$, we get ${\rm Im}\tau_{1}=0$ to obtain the odd generation.
Thus, we can not construct three tilted tori models.}

Let us study the case $(1)$.
Since $I_{dd*}=0, I_{db}=n$ and $I_{cc*}=0, I_{ac}=-n$, we find that 
$(n_{d}^{1},m_{d}^{1})=(n_{d}^{1},0)$ and $(n_{c}^{2},m_{c}^{2})=(n_{c}^2,0)$.
Then, we have 
\begin{equation} I_{ac}=n_{a}^{1}m_{c}^{1} \cdot (-m_{a}^{2} n_{c}^{2}) \cdot (n_{a}^{3}m_{c}^{3} - m_{a}^{3}n_{c}^{3})=-n,\end{equation}
\begin{equation} I_{ac*}=-n_{a}^{1}m_{c}^{1} \cdot (-m_{a}^{2} n_{c}^{2}) \cdot (-n_{a}^{3}m_{c}^{3} - m_{a}^{3}n_{c}^{3})=-n,\end{equation}
which reduce to $-n_{a}^{1}m_{c}^{1} \cdot m_{a}^{2} n_{c}^{2}\cdot n_{a}^{3} m_{c}^{3} =-n$ and $m_{a}^{3} n_{c}^{3} =0$.
On the other hand, the RR tadpole condition requires
\begin{equation} \sum_{x\in{a,b,c,d}} N_{x} m_{x}^{1}n_{x}^{2} m_{x}^{3}=  m_{c}^{1} \cdot n_{c}^2 \cdot m_{c}^{3}  =0. \end{equation}
That leads to $n=0$, and we can not obtain non-trivial solutions.
Similarly, we can show that the case $(2)$ does not lead to non-trivial solutions.

Next, let us discuss the case $(3)$.
In this case, all the possible $D6_{c,d}$-brane configurations are classified as follows, 
\begin{description}
\item[(3a)]$(n_{c}^{2},m_{c}^{2})=(n_{c}^{2},0), (n_{d}^{1},m_{d}^{1})=(n_{d}^{1},0)$,
\item[(3b)]$(n_{c}^{2},m_{c}^{2})=(n_{c}^{2},0), (n_{d}^{3},m_{d}^{3})=(n_{d}^{3},0)$,
\item[(3c)]$(n_{c}^{3},m_{c}^{3})=(n_{c}^{3},0), (n_{d}^{1},m_{d}^{1})=(n_{d}^{1},0)$.
\end{description}

In the case $(3a)$, the condition on intersecting numbers (\ref{eq:int_n}) 
and the tadpole conditions require 
\begin{eqnarray}
-n_{a}^{1}m_{c}^{1} \cdot m_{a}^{2} n_{c}^{2}\cdot n_{a}^{3} m_{c}^{3} =n, \qquad 
n_{a}^{1} m_{a}^{2} n_{a}^{3}=n_{d}^{1} m_{d}^{2} n_{d}^{3},\qquad 
m_{a}^{3}n_{c}^{3}=0, \qquad \nonumber \\
m_{d}^{3}n_{c}^{3}=0,\qquad 
m_{b}^{1} n_{b}^{2}m_{b}^{3} +m_{c}^{1} n_{c}^{2}m_{c}^{3}=0, \qquad 
3n_{a}^{1}m_{a}^{2}m_{a}^{3}+n_{d}^{1}m_{d}^{2} m_{d}^{3}=0.
\end{eqnarray}
These results  are shown in Table \ref{tab:2til1-n}.
For $n=3$, this result leads to the models in Table  \ref{tab:Sp(2)SM}.

\begin{table}[ht]
\begin{center}
\begin{tabular}{|l||ccc|}\hline
D brane �& $T^2_1$ & $T^2_2$ &  $T^2_3$ \\
\hline \hline
a &  ($n_{a}^{1},0$)  &  ($n_{a}^2,m_{a}^2$)  & ($-\rho/n_{a}^{1} m_{a}^{2},m_{a}^{3}$)\\
b &  (0,$ m_{b}^{1}$)  &  ($n_{b}^{2},0$)  & ($0, n/\rho m_{b}^{1} n_{b}^{2}$)\\
c &  ($n_{c}^1,m_{c}^{1}$)  &  ($n_{c}^{2} ,0$)  & ($n_{c}^{3},-n/\rho m_{c}^1 n_{c}^{2}$)\\
d &  ($n_{d}^1,0$)  &  ($n_{d}^2,m_{d}^2$)  & ($-\rho/n_{d}^{1} m_{d}^{2},m_{d}^{3}$)\\ \hline
\end{tabular}
\end{center}
\caption{The SM-like models with $n$ generations. $n^i_k,m^i_k$ are integer parameters satisfying 
$m_{a}^{3} n_{c}^{3} =0$,$m_{d}^{3} n_{c}^{3} =0$ and 
$3n_{a}^{1}m_{a}^{2} m_{a}^{3} +n_{d}^{1}m_{d}^{2} m_{d}^{3} =0$.
$\rho$ is a divisor of n.
To get the correct gauge symmetry, $(n_x^i,m_x^i-{\rm Im}\tau_i n_x^i)$ have to be coprime.}
\label{tab:2til1-n}
\end{table}

Similarly, we can discuss the other cases.
As a result, we find that the case  $(3b)$ is allowed only for $n=$ even, 
and the case $(3c)$  does not have non-trivial solutions.

As a result, only the case ($3a)$ has non-trivial solutions with $n=3$, and  
they are the models with the SM chiral matter fields 
as shown in  Table  \ref{tab:Sp(2)SM} for $n=3$.
However, at this stage the gauge symmetry of our models  is 
$SU(3)\times SU(2) \times U(1)_a \times U(1)_c \times U(1)_d$.
The hypercharge $U(1)_Y$ corresponds to the 
linear combination, $\frac16 U(1)_a - \frac12 U(1)_c - \frac12 U(1)_d$.
We require the other two extra $U(1)$ gauge bosons to become massive by couplings with $B^k_2$.
Here, we examine these couplings.
As the basis $[\alpha_k]$,  we set 
\begin{eqnarray}
{}[ \alpha_1 ] = (1,0)\times (0,1)\times (0,1), \nonumber \\  
{}[ \alpha_2] = (0,1) \times (1,0)\times (0,1),\\
{}[\alpha_3] = (0,1)\times (0,1)\times (1,0).  \nonumber
\end{eqnarray}
Each of $B^k_2$ couples to $U(1)$s as 
\begin{eqnarray}
 B_2^1 \wedge  m_c^1 n_{c}^{2} n_c^3 F_c, \nonumber \\
-\rho B_2^2 \wedge  (3F_a + F_d), \\
B_2^3 \wedge (3n_a^1 n_a^2m_a^3 F_a - \frac{n_c^1 n}{\rho m_c^1} F_c + n_d^1 n_d^2 m_d^3 F_d) 
\nonumber.
\end{eqnarray}
The condition for the $U(1)_Y$ gauge boson to remain massless is given by 
\begin{equation}
n_c^3 =0,  \qquad 
\redtwo{\frac{1}{2}}n_a^1 n_a^2m_a^3 \redtwo{+\frac{1}{2}}\frac{n_c^1 n}{\rho m_c^1} \redtwo{-\frac{1}{2}} n_d^1 n_d^2 m_d^3=0.
\end{equation}
If $n_c^1$ is not zero, the extra gauge bosons become massive.

\section{One-loop corrections to gauge couplings}

In section \ref{sec:gauge}, assuming that the extra fields are all sufficiently massive, we evaluated the gauge couplings at a high energy scale by using  the renormalization group equations in the SM.
In this section, we examine the validity of this assumption.
There are two types of extra fields which can be light compared to $M_s$.
One is given by the superpartners of the SM fields and the other corresponds to the Kaluza Klein(KK) and winding modes of open strings stretching between parallel D-branes.

\subsection{The superpartners of the SM fields}

In toroidal D-brane models, a single stack of D-branes preserves $\mathcal{N}=4$ supersymmetry in four dimensional field theory.
The gauge bosons have supersymmetric partners: four gauginos and three complex scalars.
However, these supersymmetries are broken by D-brane intersections and the superpartners obtain masses $M_a$ by the loop correction\cite{Ibanez:2001nd}
\begin{equation}
M_a=\frac{g_a^2}{(4\pi)^2} \frac{(1-r)}{\sqrt{1-\theta_i}} M_s,
\end{equation}
where $\theta_i$ denotes the corresponding angle of three complex scalars and $r$ is the supersymmetry breaking parameter $r=(\theta_1+\theta_2+\theta_3)/2$.
Without fine tuning, these parameters are of $\mathcal{O}(1)$ and the $g_a^2/(4\pi)^2$ are of $\mathcal{O}(10^{-2})$, therefore we obtain $M_a/M_s=\mathcal{O}(10^{-2})$.
The threshold correction due to the superpartners to $SU(N)$ gauge coupling $\alpha_a^{-1}(M_s)$ is
\begin{equation}
\Delta_f= -\frac{1}{4\pi}\frac{2N}{3}  {\rm log}\frac{M_s^2}{M_a^2}
\end{equation}
for adjoint fermions and 
\begin{equation}
\Delta_s= -\frac{1}{4\pi} \frac{N}{3} {\rm log}\frac{M_s^2}{M_a^2}
\end{equation}
for adjoint scalars.
For the $SU(3)$ case, 
These corrections are of $\mathcal{O}(1)$,
which does not significantly change our results.
The effect of the superpartners of quarks and leptons are less important than that of gauge bosons since they have mass at tree level.

If $r$ is very close to 1, $M_a$ can become very light, e.g. $M_s/M_a=\mathcal{O}(10^{10})$.
In this case, since the corrections are comparable to the tree gauge couplings,
our assumption is no longer valid .
However, even in such a  special case, the perturbative constraint in section \ref{subsec:constraint} is still valid, because these corrections make the gauge coupling bigger.
This means that the perturbative condition becomes more severe for such a case and that the red data points in Figure 1$\sim$3  
shift towards larger $(\alpha_3/\alpha_Y,\alpha_2/\alpha_Y)$, moving away from the blue data points..
These corrections only strengthen our constraint.

\subsection{The KK and winding modes }

The masses of KK and winding modes on a two dimensional torus are given by\cite{Gmeiner:2009fb,Blumenhagen:1999md}
\begin{equation}
\alpha' m_{KK}^2=\frac{\alpha'}{R_1^2 n^2 +R_2^2 m^2}\left( p+\frac{\tau}{2} \right)^2,
\end{equation}
\begin{equation}
\alpha' m_{{\rm winding}}^2=\frac{1}{\frac{\alpha'}{R_2^2} n^2 +\frac{\alpha'}{R_1^2} m^2}\left( q+\frac{\sigma}{2} \right)^2,
\end{equation}
where $\tau$ is the Wilson line in the D-brane and $\sigma$ is the displacement of two D-branes.
We do not consider the Wilson line and set $\tau$ to zero.
$\sigma$ is normalized from 0 to 1.
The masses of KK and winding modes depend on the compactification moduli parameters and winding numbers.
Here, we compute these masses 
in the explicit model shown in section \ref{subsec:explicit} and study their effects.

The masses of KK and winding modes of the SU(3) gauge boson are as follows.
\begin{equation}
\begin{split}
\alpha' m_{KK,SU(3)}^2 &=\frac{\alpha'/K_1}{10^{-2/3}}p_1^2+\frac{\alpha'/K_2}{10^{-14/9}3^2+ 10^{14/9}\left(\frac{1}{2} \right)^2}p_2^2+\frac{\alpha'/K_3}{10^{-2/3}3^2+ 10^{2/3} \left( \frac{1}{2} \right)^2}p_3^2, \\
                               &\simeq \frac{\alpha'}{K_1}4.6p_1^2+\frac{\alpha'}{K_2} \frac{1}{9.2}p_2^2+\frac{\alpha'}{K_3}\frac{1}{3.1}p_3^2,\\
\alpha' m_{{\rm winding},SU(3)}^2 &=\frac{K_1/\alpha'}{10^{2/3}}q_1^2+\frac{K_2/\alpha'}{10^{14/9}3^2+ 10^{-14/9}\left(\frac{1}{2} \right)^2}q_2^2+\frac{K_3/\alpha'}{10^{2/3}3^2+ 10^{-2/3} \left( \frac{1}{2} \right)^2}q_3^2, \\
                               &\simeq \frac{K_1}{\alpha'}\frac{1}{4.6} q_1^2+\frac{K_2}{\alpha'}\frac{1}{323}q_2^2+\frac{K_3}{\alpha'}\frac{1}{41.8}q_3^2.
\end{split}
\end{equation}
$K_i$ denotes the area of the $i$th torus, and $p_i,q_i$ denote the KK momentum and winding number in the $i$th torus.
We set $K_1 K_2 K_3=\alpha'^3$ in the previous analysis.
The masses of SU(2) gauge boson's KK and winding modes are
\begin{equation}
\label{eq:SU(2)KK}
\begin{split}
\alpha' m_{KK,SU(2)}^2 &=\frac{\alpha'/K_1}{10^{2/3}}p_1^2+\frac{\alpha'/K_2}{10^{-14/9}2^2}p_2^2+\frac{\alpha'/K_3}{ 10^{2/3}}p_3^2, \\
                                &\simeq \frac{\alpha'}{K_1}\frac{1}{4.6}p_1^2+\frac{\alpha'}{K_2} \frac{1}{0.11}p_2^2+\frac{\alpha'}{K_3}\frac{1}{4.6}p_3^2,\\
\alpha' m_{{\rm winding},SU(2)}^2 &=\frac{K_1/\alpha'}{10^{-2/3}}q_1^2+\frac{K_2/\alpha'}{10^{14/9}2^2}q_2^2+\frac{K_3/\alpha'}{ 10^{-2/3}}q_3^2, \\
                               &\simeq \frac{K_1}{\alpha'}4.6 q_1^2+\frac{K_2}{\alpha'}0.11q_2^2+\frac{K_3}{\alpha'}4.6q_3^2.
\end{split}
\end{equation}
There are also the superpartners of the KK and winding modes of gauge bosons, that is, 
4 gauginos and 3 complex scalars.
Their masses are
\begin{equation}
m^2_{{\rm superpartners},SU(N)} = m_{KK,SU(N)}^2+m_{{\rm winding},SU(N)}^2+\delta_{SU(N)},
\end{equation} 
where $\delta_{SU(N)}$ denotes the supersymmetry breaking effect.
In addition to that, there are extra  fermions having gauge charge. 
The masses of extra fermions between the D6$_a$-brane and the D6$_{d,d^*}$-brane are written as 
\begin{equation}
\begin{split}
\alpha' m_{KK,ad^{(*)}}^2 +\alpha' m_{{\rm winding},ad}^2 &=\frac{\alpha'/K_1}{10^{-2/3}}p_1^2+\frac{\alpha'/K_1}{10^{2/3}}(q_1+\sigma_{ad^{(*)}})^2,\\
                                &\simeq \frac{\alpha'}{K_1}4.6p_1^2+\frac{K_1}{\alpha'}\frac{1}{4.6}(q_1+\sigma_{ad^{(*)}})^2.\\
\end{split}
\end{equation}
These two fermions have almost the same masses of KK and winding modes.
The only difference is the distance $\sigma_{ad^{(*)}}$ and generation number. 
The ad mode has four generations and the ad$^*$ mode has eleven generations.
These are all of the  SU(3) charged KK and winding modes which can be significantly light.

The one-loop threshold correction for SU(3) gauge coupling is computed as
\begin{equation}
\begin{split}
\Delta_{SU(3)}
\simeq -\frac{1}{4\pi}11& \sum_{KK,{{\rm winding}}} {\rm log}\left(1+ \frac{\delta_{SU(3)}}{m_{KK,SU(3)}^2+m_{{\rm winding},SU(3)}^2}\right)\\
                  &+\frac{1}{4\pi}\frac{4}{3}15 \sum_{KK,{{\rm winding}}} {\rm log}\left( 1+\frac{\delta_{ad^{(*)}}}{m_{KK,ad^{(*)}}^2+m_{{\rm winding},ad^{(*)}}^2}\right),
\end{split}
\end{equation}
where $\delta_{ad^{(*)}}$ denotes a supersymmetry breaking effect.
If $K_i$ is comparable to $\alpha'$, the lightest mode is the gauge boson's winding state on $T_2^2$.
We can approximately write
\begin{equation}
\label{eq:winding}
\begin{split}
\Delta_{SU(3)} &\simeq -\frac{11}{4\pi} \sum_{p_2^2<323\frac{\alpha'}{K_2}} {\rm log}\left( 1+\frac{323 \alpha'^2 \delta_{SU(3)}}{K_2 p_2^2} \right),\\
                   &\simeq -\frac{11}{4\pi} \int_1^{M} dx {\rm log} (1+\alpha'\delta_{SU(3)} M^2/x^2),\\
                   &=-\frac{11}{4\pi} \Biggl( M{\rm log}(1+\alpha' \delta_{SU(3)}) -{\rm log}(1+\alpha' \delta_{SU(3)} M^2) \Biggr. \\
     & \left. +i\sqrt{\alpha' \delta_{SU(3)}}M\left({\rm log}\frac{1+i\sqrt{\alpha' \delta_{SU(3)}}}{1-i\sqrt{\alpha' \delta_{SU(3)}}} -{\rm log}\frac{1+i\sqrt{\alpha' \delta_{SU(3)}}M}{1-i\sqrt{\alpha' \delta_{SU(3)}}M}\right)\right),\\
                   &=-\frac{11}{4\pi}(M^2 -M)\alpha' \delta_{SU(3)} +\mathcal{O}((\alpha' \delta_{SU(3)})^2),
\end{split}
\end{equation}
where $M$ denotes $(323\alpha'/K_2)^{1/2}$.
As mentioned in the previous subsection, the supersymmetry breaking effect $\delta_{SU(3)}$ is of $(10^{-2} M_s)^2$ unless fine tuning is applied.
Then, $\Delta_{SU(3)}$ is of $\mathcal{O}(10^{-2}\alpha'/K_2)$.
Using the SM renormalization group equations, 
we obtain $1/\alpha_3'(10^{18}{\rm GeV})\sim 40$.
The above threshold correction to the gauge coupling is sufficiently small if $K_2/\alpha'<10^{-2}$.

In the $SU(2)$ sector, considering (\ref{eq:SU(2)KK}), there are not so many $SU(2)$ charged KK modes and winding modes lighter than $M_s$.
The correction of $SU(2)$ would be smaller than of $SU(3)$.
This holds true for the $U(1)_Y$ gauge coupling, too.

To summarize, in the model shown in section \ref{subsec:explicit}, we conclude the one-loop threshold corrections due to massive modes are irrelevant if the area of the torus $K_i/\alpha'$ is of $\mathcal{O}(1)$.
Other models may lead to similar behavior.

\end{document}